\documentclass{aa}  

\usepackage{graphicx}
\usepackage{txfonts}
\usepackage{amsmath}	
\usepackage{amssymb}	
\usepackage{multirow}
\usepackage{float}
\usepackage{siunitx}
\usepackage{booktabs}
\usepackage{color,soul}
\usepackage[utf8]{inputenc}
\usepackage[para]{threeparttable}

\graphicspath{ {./../Astronomy_Astrophysic/} }

\newcommand{\Msun}{M$_{\hbox{$\odot$}}$}  
\newcommand{\cmc}{cm$^{-3}$}              
\newcommand{\kms}{km\,s$^{-1}$}           
\newcommand{\mum}{$\mu$m} 
\newcommand{\Lco}{K\,\kms\,pc$^{2}$}
\newcommand{\Xco}{\Msun\,(K\,\kms\,pc$^{2}$)$^{-1}$}
\newcommand{\lratio}{$\text{CO}(2-1)/\text{CO}(1-0)$}
\newcommand{\hratio}{$\text{CO}(3-2)/\text{CO}(1-0)$}

\begin{document}

   \title{SuperCAM CO($3-2$) APEX survey at $6$ pc resolution in the\\ Small Magellanic Clouds}


      \author{H. P. Salda\~no\inst{1,}\inst{2},
          M. Rubio\inst{3},
          A. D. Bolatto\inst{4},
          K. Sandstrom\inst{5},
          B. J. Swift\inst{6},
          C. Verdugo\inst{7},
          K. Jameson\inst{8},
          C. K. Walker\inst{6},
          C. Kulesa\inst{6},
          J. Spilker\inst{9},
          P. Bergman\inst{10},
          G. A. Salazar\inst{1,2}
          }
          
   \institute{Instituto de Investigaciones en Energ\'ia no Convencional, Universidad Nacional de Salta, C.P.: 4400, Salta, Argentina\\
              \email{hpablohugo@gmail.com}
         \and
             Consejo Nacional de Investigaciones Cient\'ificas y T\'ecnicas (CONICET), Godoy Cruz 2290, CABA, CPC 1425FQB, Argentina
         \and
             Departamento de Astronom\'ia, Universidad de Chile, Casilla 36-D, Santiago, Chile
        \and
             Department of Astronomy and Joint Space-Science Institute, University of Maryland, College Park, MD 20742, USA
        \and 
             Center for Astrophysics \& Space Sciences, Department of Physics, University of California, San Diego, 9500 Gilman Drive, San Diego, CA 92093, USA
        \and 
            Department of Astronomy and Steward Observatory, University of Arizona, 85719, Tucson, AZ, USA
        \and
            Joint ALMA Observatory (JAO), Alonso de C\'ordova 3107, Vitacura, Santiago de Chile
        \and Owens Valley Radio Observatory, California Institute of Technology, Big Pine, CA 93513, USA,
        \and
        Department of Physics and Astronomy and George P. and Cynthia Woods Mitchell Institute for Fundamental Physics and Astronomy, Texas A\&M University, College Station, TX, USA
        \and Onsala Space Observatory, Department of Space, Earth and Environment, Chalmers University of Technology, SE-439 92, Onsala, Sweden
        }
        

 
  \abstract
   {The Small Magellanic Cloud (SMC) is an ideal laboratory to study the properties of star-forming regions due to its low metallicity, which affects the molecular gas abundance. However, a small number of molecular gas surveys of the entire galaxy were made in the last few years, limiting the measure of the interstellar medium (ISM) properties in a homogeneous manner.}
   {We present the CO($3-2$) APEX survey at $6$ pc resolution of the bar of the SMC, observed with the SuperCAM receiver attached to the APEX telescope. This high-resolution survey allowed us to study some properties of the ISM and the identification of CO clouds in the innermost part of the H$_2$ envelopes.}
   {We aboard the CO analysis in the SMC-Bar comparing the CO($3-2$) survey with that of the CO($2-1$) of similar resolution. We study the CO($3-2$)-to-CO($2-1$) ratio ($R_{32}$) that is very sensitive to the environment properties (e.g., star-forming regions). We analyzed the correlation of this ratio with observational quantities that trace the star formation as the local CO emission, the Spitzer color $[70/160]$, and the total IR surface brightness measured from the Spitzer and Herschel bands. For the identification of the CO($3-2$) clouds, we used the CPROPS algorithm, which allowed us to measure the physical properties of the clouds. We analyzed the scaling relationships of such physical properties.}
   {We obtained an $R_{32}$ of $0.65$ as a median value for the SMC, with a standard deviation of $0.3$. We found that $R_{32}$ varies from region to region, depending on the star formation activity. In regions dominated by HII and photo-dissociated regions (e.g., N22, N66), $R_{32}$ tends to be higher than the median values. Meanwhile, lower values were found toward quiescent clouds. We also found that $R_{32}$ correlates positively with the IR color $[70/160]$ and the total IR surface brightness. This finding indicates that $R_{32}$ increases with environmental properties like the dust temperature, the total gas density, and the radiation field. We have identified $225$ molecular clouds with sizes $R > 1.5$ pc and signal-to-noise (S/N) ratio $> 3$ and only $17$ well-resolved CO($3-2$) clouds increasing the  S/N ratio to $\gtrsim 5$. These $17$ clouds follow consistent scaling relationships to the inner Milky Way clouds but with some departure. The CO($3-2$) tends to be less turbulent and less luminous than the inner Milky Way clouds of similar size. Finally, we estimated a median virial-based CO-to-H$_2$ conversion factor of $12.6_{-7}^{+10}$ \Xco\ for the total sample.}
   {}

                
    \titlerunning{SuperCAM CO($3-2$) survey in the SMC}
    \authorrunning{H. P. Salda\~no, et al.}

    \maketitle
%

\section{Introduction}

The Small Magellanic Cloud (SMC), the nearest low metallicity galaxy to the Milky Way \cite[$\sim 60$ kpc,][]{Hilditch_2005MNRAS_357_304H}, with a low metal abundance \citep[$Z \approx  0.2\,Z_{\hbox{$\odot$}}$,][]{Russell_1992ApJ_384_508R} and high Gas-to-Dust ratio \citep[GDR $\sim 2000$,][]{Roman_Duval_2017ApJ_841_72R}, has presented a  challenge in the study of the molecular component of the interstellar medium (ISM) and its association with star-forming regions. The low dust content prevents protection from the high ultra-violet (UV) radiation field allowing a strong dissociation of most of the molecular gas at low $A_V$, specifically carbon monoxide (CO), the common tracer of H$_2$ gas.

Theoretical models \citep{Wolfire_2010_ApJ_716_1191W,Seifried_2020_MNRAS_492_1465S} show that due to the strong H$_{2}$ self-shielding, the H$_2$ molecule starts to form in low hydrogen nuclei density ($n_{\text{H}} = 10 - 10^{3}$ \cmc) regions where $A_V$ takes values from $0.2$ to $1.5$ mag. These molecular H$_2$ regions, also containing neutral (C) and ionized carbon (C$^{+}$) but without CO gas, are frequently embedded within strong photodissociation regions (PDRs) well traced by the [CII] emission at 158 \mum\ \citep[see also][]{Requena_Torres_2016_AA_589A_28R,Pineda_2017_ApJ_839_107P,Jameson_2018_ApJ_853_111J}. These molecular envelopes where CO is not present are known as "dark molecular gas". At higher densities ($n_{\text{H}} > 300$), where $A_V \gtrsim 1.5$ and the temperatures are below $\sim$ 50 K, the carbon is present as CO and becomes the most abundant molecular component after the H$_2$ molecule ("bright CO gas"). Generally, in normal conditions such as those present in the inner Milky Way, the total molecular gas traced by CO makes a $60\%$ to $80\%$ contribution \citep{Wolfire_2010_ApJ_716_1191W}. However, at very low metallicities, like that of the SMC, the molecular and atomic carbon distributions change considerably to a degree that the CO filling fraction can be reduced several orders of magnitude to that of their Galactic counterparts \citep{Bolatto_2013ARA&A_51,Kalberla_2020_AA_639A_26K}. The CO gas only will trace less than $30$\% \citep{Pineda_2017_ApJ_839_107P,Jameson_2018_ApJ_853_111J} of the total H$_2$ gas in the innermost part of the molecular clouds \citep[see also][]{Oliveira_2011MNRAS_411L_36O}. The bright CO gas fraction may also be reduced in magnetized molecular clouds \citep{Seifried_2020_MNRAS_492_1465S}. These characteristics strongly difficult the analysis of the molecular gas in the SMC, leaving the studies only to small CO structures embedded in larger H$_2$ clouds.

However, despite that CO becomes a poorer indicator of the total H$_2$ gas in low metallicity galaxies, CO observations at different transitions \citep{Bolatto_2008ApJ_686,Rubio_2000AA_359_1139R,Muller_2014_PASJ_66_4M,Jameson_2018_ApJ_853_111J,Oliveira_2019MNRAS_490_3909O,Saldano_2023AA_672A_153S} has allowed analyzing the dynamical state and physical conditions of cold dense molecular clouds where the star formation is expected to take place. The first rotational transitions of CO ($J=1-0$ and $J=2-1$) trace the cold bright CO gas of low density because of its low excitation energies ($E_{u}/k = 5.5-16.6$ K) and lower critical densities ($10^{3} \lesssim n_{crit} \lesssim 10^{4}$ \cmc)\footnote{ http://th.nao.ac.jp/MEMBER/tomisaka/research\_resources/catalog.html}. Higher-J transitions like CO($J=3-2$), are required to study warmer and denser regions of molecular clouds, as this transition has an excitation energy of $E_{u}/k = 33$ K and a critical density of $n_{crit} \sim 7\times10^{4}$ \cmc. Combining high-J transitions with lower-J transitions has allowed a better constrain of the temperatures and densities within the molecular clouds.

Multi-transitional models \citep{Penialoza_2018_MNRAS_475_1508P} and observations of the CO emission in different galaxies \citep{denBrok_2021_MNRAS_504_3221D,Leroy_2022ApJ_927_149L} have shown that CO line ratios of different transitions are very sensitive to the physical conditions of the environment, such as the dust temperature, column density, interstellar radiation field (ISRF),  cosmic ray, and star formation rate (SFR). In the SMC, \cite{Bolatto_2003ApJ_595_167B} found a CO(2$-$1)/CO(1$-$0) integrated line brightness ratio $\gtrsim 2$ (at $43''$ resolution) in the N83/N84 molecular cloud. The high ratio is associated with an expanding molecular shell which coincides with the NGC 456 stellar association and a supernova remnant. \cite{Nikolic_2007_AA_471_561N}, performed a multi-transitional study of CO in six different regions of the SMC, which included quiescent clouds with no signs of star formation activity and others clouds with star formation activity. In the first case, low \lratio\ and \hratio\ ratios (at $45''$ resolution) with values $\lesssim 0.5$ were found in a warm ($50$ K) and relatively dense gas ($n_{\text{H}_2} = 700$ \cmc) but quiescent molecular cloud SMCB1\#1. However, higher \lratio\ and \hratio\ ratios ($\gtrsim 1)$ were found in other regions associated with prominent star formation, like N12, N27, N66, and N83. In these active regions, the observed line ratios were reproduced by a two-component model comprised of a cold dense component ($T_{k} = 10-150$ K, $n_{\text{H}_2} = 10^{4}-10^{5}$ \cmc) and a hot tenuous component ($T_{k} = 100-300$ K, $n_{\text{H}_2} = 10^{2}-10^{3}$ \cmc).

In the Magellanic Bridge (MB), \cite{Muller_2014_PASJ_66_4M} made a study of the \hratio\ in several clouds at $22''$ resolution. A much higher \hratio\ ratio of $\sim 2.5$ was found towards the cloud MB-B, which is associated with recent or current star formation. However, this high ratio is not seen in other sites of the Magellanic Bridge, like the clouds MB-A and MB-C, which show \hratio\ ratio $\lesssim 1$. Such clouds may be remnants of a past period of star formation.

Toward the Large Magellanic Clouds (LMC), high \hratio\ ratios between $\sim 1-2$  were found in warm ($60-80$ K) and dense ($n_{\text{H}_2} > 10^{3}$ \cmc) clouds which correlate with the 24 \mum\ and H$\alpha$ peak emission \citep{Minamidani_2008_ApJS_175,Mizuno_2010_PASJ_62_51M}, while lower rations were found in cooler and lower dense clouds poorly associated with star-forming indicators \citep[see also][]{Celis_2019_AA_628A_96C}. In the starburst-like galaxy NGC 1140, \cite{Hunt_2017_AA_606A_99H} found that high $R_{31} \sim 2$ is an indication of high H$_2$ density ($\sim 10^{6}$ \cmc) in cool clouds ($\lesssim 20$ K) and also a sign of somewhat excited, optically thin gas. 

In this paper, we present a CO($3-2$) survey of the main body of the SMC (SMC-Bar) performed with SuperCAM attached to the APEX telescope at 6 pc resolution. In combination with the CO($2-1$) survey of the SMC \citep{Saldano_2023AA_672A_153S} we have determined the CO($3-2$)-to-CO($2-1$) ratio of the SMC and studied this ratio for clouds located in different environments within the galaxy. In addition, we have derived and analyzed the physical properties of the CO($3-1$) clouds and studied their scaling relations. Our results will impact the study of external galaxies where CO(3-2) emission is being used to determine the mass and properties of the molecular gas. In the following section (Sec. \ref{sec:data_and_reduction}), we detail the reduction of the SuperCAM data and we show the complementary data used for the analysis. In Sect. \ref{sec:CO_ratio_method}, we show the methodology used to estimate the CO($3-2$)-to-CO($2-1$) ratio in different parts of the SMC, and in Sect. \ref{sec:clouds_cprops}, we explain the cloud identification by the CPROPS algorithm. In Sect. \ref{sec:results}, the main results are shown, while in Sect. \ref{sec:discussion} the discussion of the main results are presented. Finally, In Sect. \ref{sec:conclusions}, we present the conclusion and final remarks.

   \begin{figure}
   \centering
   \includegraphics[width=1.06\linewidth]{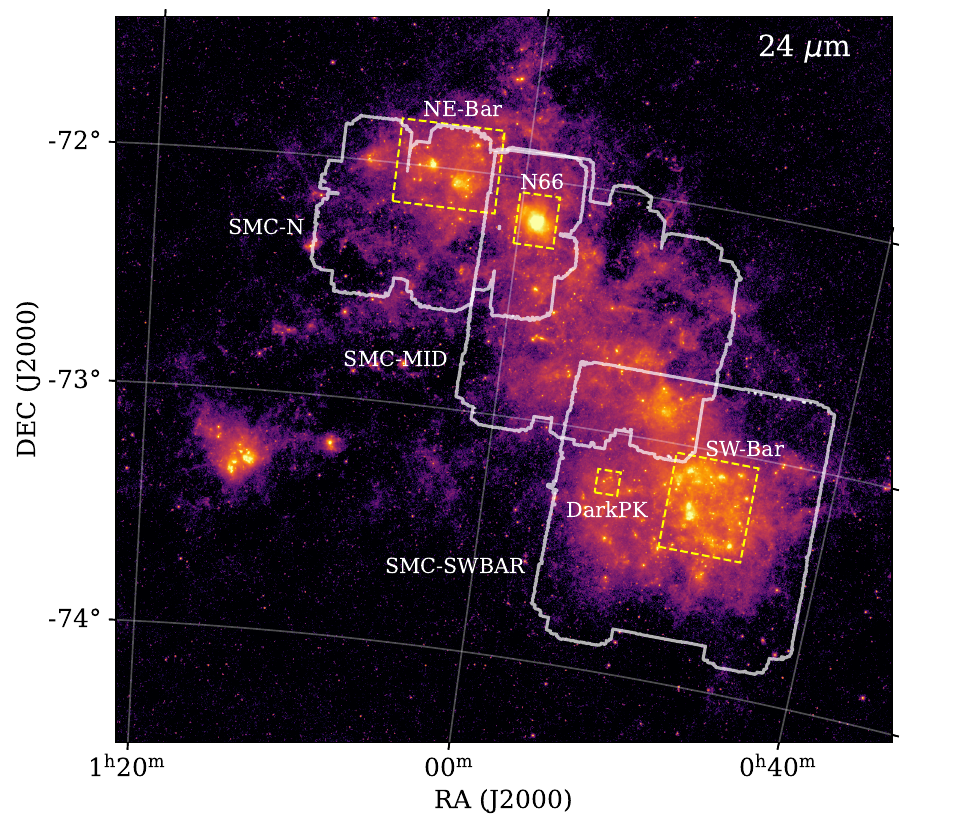}
         \caption{SuperCAM observations of the SMC delineated by white contours. These contours are superimposed on the 24 \mum\ SAGE-SMC image \citep{Gordon_2011_AJ_142_102G}. The yellow boxes show the CO($2-1$) APEX maps \citep{Saldano_2023AA_672A_153S}.}
         \label{fig:smc_supercam}
   \end{figure}


\section{Data and Reduction}
\label{sec:data_and_reduction}

\subsection{CO($3-2$) Observations}
\label{sec:CO_data}

The SMC survey was performed using the SuperCAM multi-pixel focal plane array of superconducting mixers attached to the 12m Atacama Pathfinder EXperiment (APEX\footnote{APEX is a collaboration between the Max-Planck-Institut für Radioastronomie, the European Southern Observatory, and the Onsala Space Observatory. Swedish observations on APEX are supported through Swedish Research Council grant no. 2017-00648.http://www.apex-telescope.org/}) \citep{Gusten_2006_AA_454L_13G} telescope, located in Llano de Chajnantor, near San Pedro de Atacama, Northern Chile. The galaxy was mapped in the CO(3$-$2) line during the period the visiting instrument SuperCAM was deployed. The observations were done on two dates, 2014 December 08, and 2015 May 16, 2015 (PI. M. Rubio, Project C-095.F-9707B, C-094.F-9306A, PI. A. Bolatto O-094.F-9306A).  

SuperCAM consists of a 64-pixel $345$ GHz heterodyne imaging spectrometer designed to operate in the astrophysically important $870\,\mu$m atmospheric window \citep{Walker_2005stt_conf_427W}. This camera was built by the Steward Observatory Radio Astronomy Laboratory (SORAL\footnote{http://soral.as.arizona.edu/Supercam/Welcome.html}), a project belonging to the University of Arizona, and is by far the largest such array used to perform very large surveys of the sky. For instance, it was used to map a $2.7$ square degree area of the Orion molecular cloud complex \citep{Stanke_2022AA_658A_178S}. For more technical details of the array receiver see \cite{Groppi_2010stt_conf_368G} and \cite{Kloosterman_2012SPIE_8452E_04K}. 

\begin{figure*}
\centering
\includegraphics[width=0.6\linewidth]{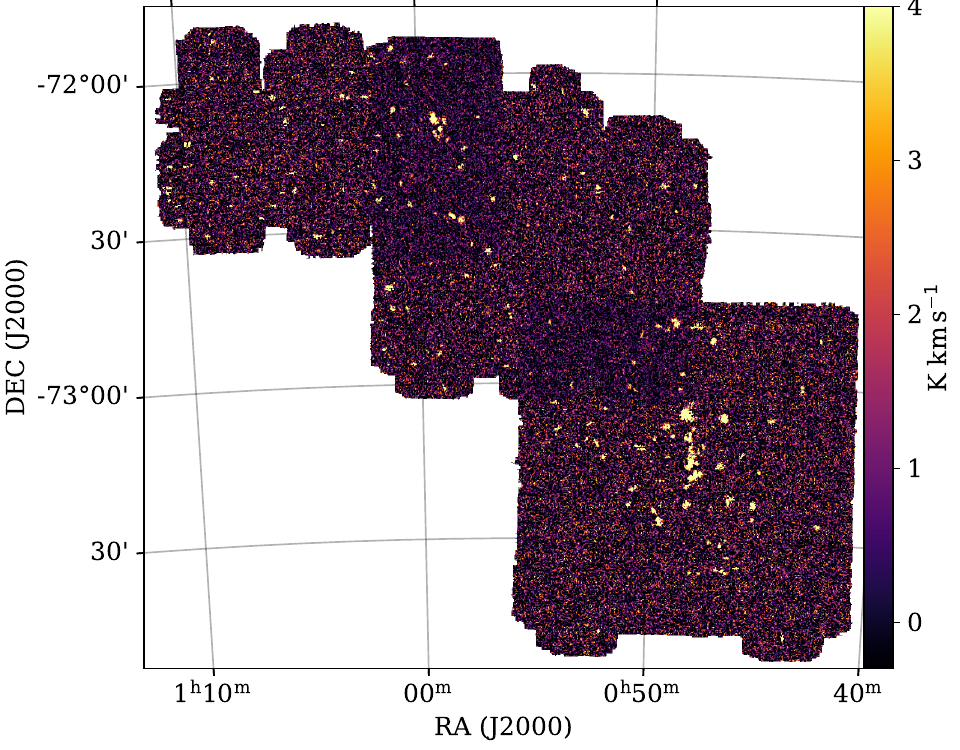}
   \caption{CO($3-2$) integrated map of the SMC-BAR observed by SuperCAM. The integration was done within the velocity range of clouds with high S/N ratios ($> 3$). We included an artificial homogeneous background noise in the map to enhance the appearance of the detected clouds. The real background noise is shown in Fig. \ref{fig:smc_rms}.} 
\label{fig:smc_integrated}
\end{figure*}

\begin{table*}
\centering
\caption{Characteristics of the SuperCAM CO($3-2$) observations in the SMC}
\begin{threeparttable}
\begin{tabular}{lcccccc}
\hline 
\hline
SMC   &     R.A.   &     DEC       &         FoV        & HPBW & $\Delta V$\tnote{a} & $T_{rms}$\tnote{b} \\
 &(hh:mm:ss.s)& (dd:mm:ss.s)  &  ($^{\circ}\times^{\circ}$) &  (arcsec) &    (\kms)    &     (K)     \\
\hline
SWBAR & 00:48:18.8 & $-$73:18:05.5 & 1.1$\,\times\,$1.2 &  20  &    0.4     &   1.0  \\
MID   & 00:54:48.5 & $-$72:28:27.9 & 1.1$\,\times\,$1.2 &  20  &    0.4     &   0.8  \\
N     & 01:03:43.0 & $-$72:12:36.5 & 1.1$\,\times\,$0.8 &  20  &    0.4     &    0.9 \\
\hline
\end{tabular}
\label{tab:observations}
\begin{tablenotes}
\item[a] The original cubes (with $\Delta V = 0.23$ \kms) were re-sampled to a spectral resolution of $0.4$ \kms\ to increase the signal-to-noise ratio.\\
\item[b] The median $T_{rms}$ is calculated in $0.4$~km\,s$^{-1}$ spectral resolution.   
\end{tablenotes}
\end{threeparttable}
\end{table*}

The SMC was mapped in multiple overlapping regions of size $\sim 20'\times10'$ using the on-the-fly (OTF) mode. These maps were later combined to improve the signal-to-noise ratio (S/N) of different regions of the galaxy. This process generated three large maps, SMC-SWBAR, SMC-MID, and SMC-N, that cover most of the SMC-BAR as is shown in Figure \ref{fig:smc_supercam}. In the figure, the white contours show the observed SuperCAM area superimposed on the  24 \mum\ emission image. The figure also shows the areas of CO(2-1) mapping \citep{Saldano_2023AA_672A_153S}. The reduction of the OTF maps, as well as the combination of all of them, were made using an automatic procedure following a set of GILDAS/CLASS and Python algorithms developed by the SORAL team, which include the baseline fitting to each spectrum in the cubes using a linear polynomial fit. As a reference for the baseline fitting, we used the NANTEN CO($1-0$) data cube \citep[]{Mizuno_2001PASJ_53L_45M,Mizuno_2009IAUS_256_203M}.

The antenna temperature, $T_{A}$, was transformed to brightness temperature of the main beam ($T_{mb} = T_{A}/\eta_{mb}$) using the telescope beam efficiency $\eta_{mb} =  0.38$. The final reduction produced CO($3-2$) cubes with a spatial resolution of HPBW $= 20$" ($\sim 6$ pc at the SMC distance) and a spectral resolution of $0.233$ \kms. To further increase the signal-to-noise ratio of the cubes, we re-sampled the velocity axis to a spectral resolution of 0.4 \kms. The median intensity $rms$ of the mapped region is $\sim 1$ K. The final SMC-SWBAR, SMC-MID cubes have sizes of $\sim 1^{\circ}.1\times1^{\circ}.2$, while the SMC-N cube has a size of $\sim1^{\circ}.1\times0^{\circ}.8$. 

To obtain the final SMC-Bar CO(3-2) cube, we performed a mosaicing with the SMC-SWBAR, SMC-MID and SMC-N cubes. We show the integrated CO($3-2$) emission of the SMC- Bar in Fig. \ref{fig:smc_integrated}. The integration was performed within the velocity range of the clouds with S/N $> 3$. We included an artificial homogeneous background noise in the map to enhance the appearance of the detected clouds. In Appendix \ref{app:rms_emission}, we show the rms map of the CO($3-2$) emission.

\subsection{Complementary Data}

We used the APEX SMC CO($2-1$) survey reported by \cite{Saldano_2023AA_672A_153S}. The APEX CO($2-1$) survey has a spatial and spectral resolution of $30''$ ($\sim 9$ pc at the SMC distance) and $\Delta V = 0.25$ \kms, respectively, and covers a subset of regions in the SMC, namely SW-Bar, NE-Bar, N66 and DarkPK, which are shown in Figure \ref{fig:smc_supercam} by yellow boxes. The CO($2-1$) $rms$ intensity of the regions vary between $\sim 0.1 - 0.3$ K.

We also used the Infrared Array Camera (IRAC) $8$ \mum\ image and the Multiband Imaging Photometer (MIPS) $24$, $70$, and $160~\mu$m images from the Spitzer Survey "Surveying the Agents of Galaxy Evolution" \citep[SAGE;][]{Gordon_2011_AJ_142_102G}). We also included for the analysis the HERITAGE Herschel data in three bands from $100$ to $250~\mu$m from \cite{Gordon_2014_ApJ_797_85G} and the H$\alpha$ image from the "Magellanic Cloud Emission-Line Survey"  \citep[MCELS,][]{Winkler_2015ASPC_491_343W}.

\begin{figure*}
    \centering
    \includegraphics[width=0.7\linewidth]{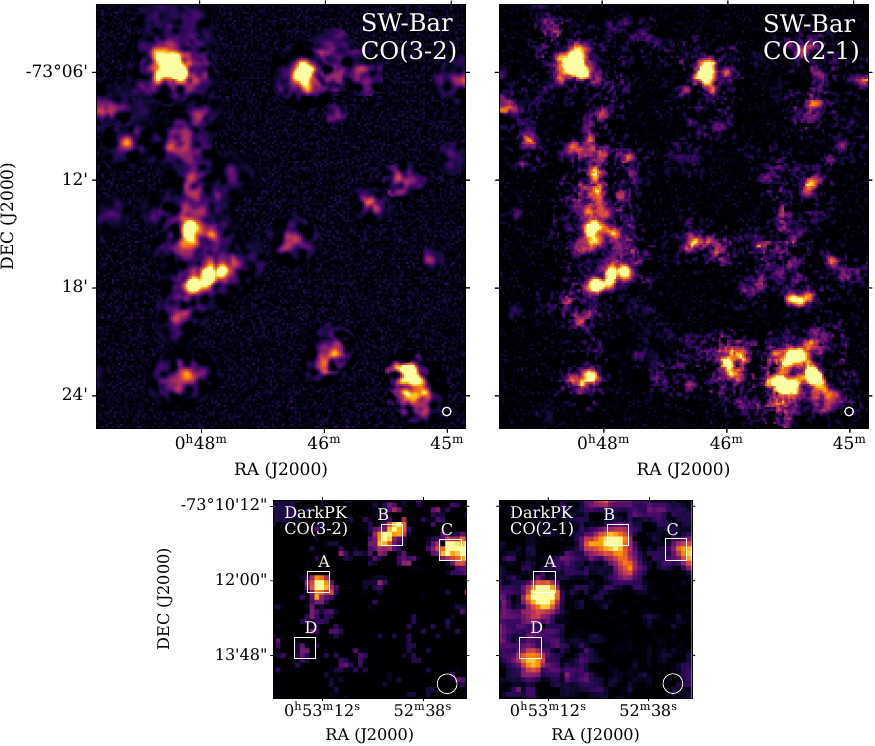}
    \caption{Integrated CO emission in the SW-Bar and DarkPK. All maps were obtained with the moment-masked method. The left panels correspond to the transition $J=3-2$, while the right panels show the CO emission in the transition $J=2-1$. The spatial resolution of $30''$ is indicated in the bottom right corner of all maps. The upper panels were used to determine $R_{32}$ pixel-by-pixel. In the DarkPK regions, the boxes A, B, C, and D indicate the positions that showed reliable CO emission in both transitions to estimate $R_{32}$.}
    \label{fig:integrated_SWBar}
\end{figure*}

\begin{figure*}
    \centering
    \includegraphics[width=0.7\linewidth]{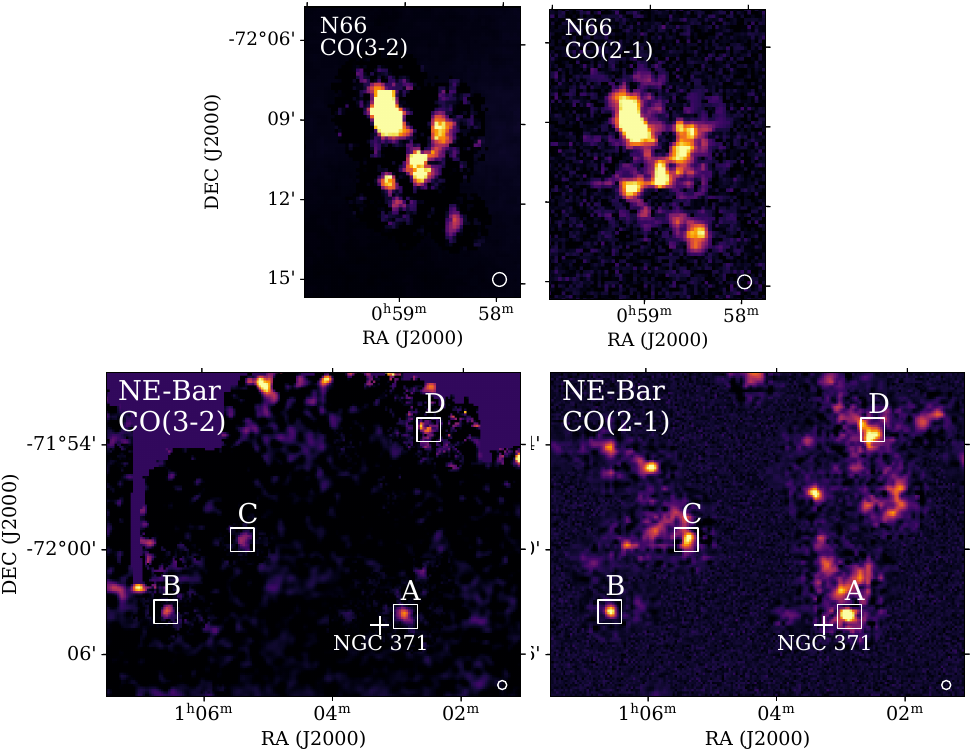}
    \caption{Similar of Fig. \ref{fig:integrated_SWBar} but for N66 and NE-Bar. The four white boxes in the NE-Bar region indicate the only positions where we found reliable CO emission in the $J=3-2$ transition.}
    \label{fig:integrated_NEBar}
\end{figure*}

\section{The CO($3-2$)-to-CO($2-1$) ratio}
\label{sec:CO_ratio_method}

We are interested in quantifying the $R_{32} =$ [CO($3-2$)/CO($2-1$)] ratio in the SMC. This ratio can be determined only towards the common areas in both transitions as the CO(2-1) maps do not cover completely the CO(3-2) map, namely SW-Bar, DarkPK, N66, and NE-Bar regions as shown in Fig. \ref{fig:smc_supercam}. The SW-bar and N66 CO($3-2$) maps show high S/N ratios so we determined $R_{32}$ using a  pixel-by-pixel method. The first step in this procedure was to convolve the CO($3-2$) data cubes to the CO($2-1$) spatial resolution of $30''$ by using a Gaussian beam of $\sim 22''$. After convolution, both the CO($3-2$) and CO($2-1$) data cubes were resampled to a common spatial grid. Then, we integrated the data cubes in both transitions using the moment masking method \citep{Dame_2011arXiv1101_1499D} and we used the following criteria for this method. We smoothed the data cube to a beam of $2$ times the original spatial resolution and in $4$ times the channel width to build the masked cube and masked the emission below $5\sigma$ rms in temperature. This mask is used to integrate the data cubes only within spectral CO lines that we consider as true brightness temperatures, avoiding spurious and unrealistic (e.g., spikes) emissions. These unrealistic emissions occur mainly in the SuperCAM data cubes that have pronounced non-uniform noise distributions (see Figure \ref{fig:smc_rms}). Finally, we calculated  $R_{32}$ pixel-by-pixel using the integrated CO($3-2$) and CO($2-1$) maps using pixels that have S/N $> 3$ in intensity.

Due to the non-uniform and much higher noise of the CO($3-2$) maps towards the DarkPK and NE-Bar, we used a different method to determine  $R_{32}$ in these two regions. Instead of a pixel-by-pixel approach, we used the CO integrated spectrum towards the strongest clouds in the two transitions identified in each region to obtain $R_{32}$ (see Figs. \ref{fig:integrated_SWBar} and \ref{fig:integrated_NEBar}). We extracted the CO($3-2$) and CO($2-1$) integrated spectra at the peak of the cloud from the convolved cubes at $30''$ resolution. The integrated emissions ($I_{\rm CO}$) in both transitions were calculated using the zeroth-order moment method between the total width of the CO lines, and finally, $R_{32}$ is calculated as the ratio between both integrated intensities.
  
\section{Cloud Identification}
\label{sec:clouds_cprops}

In addition to determining the $R_{32}$ ratio, we also identified the individual molecular clouds in CO($3-2$) at $6$ pc resolution. For the identification, we used the CPROPS algorithm \citep{Rosolowsky_2006PASP_118_590} in the three SuperCAM CO($3-2$) maps using the following criteria in the algorithm: clouds with sizes larger than the spatial resolution ($20$"), FWHM greater than $3$ channels in velocity ($> 1.2$ \kms) and $2$ sigma above the noise level. So we fixed the CPROPS input parameters: MINAREA$=1$ and MINCHAN$=3$. The THRESHOLD input parameter \citep[the cut in intensity to define "islands", see][]{Rosolowsky_2006PASP_118_590} was fixed to $2$. The EDGE input parameter was first set to $1.5$ to $1.0$ to extend the wing of the "islands". For EDGE $=1.5$, only some tens clouds were identified with S/N $> 3$ in all three cubes. We changed to EDGE $= 1$, which extends the wing of the "islands" and we found many more but weaker clouds, some of which could be false CO emissions (e.g., spikes). To discriminate between these weaker true and false clouds, we extracted the spectrum of the cloud integrating on the area of the cloud as defined by its size and selected only those clouds that have an integrated spectral line with a velocity width larger than 3 adjacent channels ($\simeq 1.2$ \kms) and temperatures $\gtrsim 3\times$rms in the channels. We considered the clouds identified by these criteria as confirmed clouds.

The properties of the CO($3-2$) clouds such as the radius ($R$ in pc), velocity dispersion ($\sigma_{\upsilon}$ in \kms), CO flux ($F_{\rm CO}$ in K\,\kms) given by CPROPS by mean of the moment method in the position-position-velocity data cube are summarized in Table \ref{tab:cprops_parameters_SMC}. In col. 1, we listed the identification (ID) number of the clouds with high S/N ($\gtrsim 5$). In col. 2 and 3, the Right Ascension and Declination in J$(2000)$ are indicated. In col. 4 to 10, we listed the radius ($R$), the {\it local-standard of rest} velocity ($V_{\rm lsr}$), the velocity dispersion ($\sigma_{\upsilon}$), the peak temperature ($T_{\text{peak}}$), the integrated CO($3-2$) intensity ($I_{\text{CO(3-2)}}$), the CO($3-2$) luminosity ($L_{\text{CO(3-2)}}$), and the virial mass ($M_{\text{vir}}$). The ID numbers in Table \ref{tab:cprops_parameters_SMC} refer to their number on the online version table, which includes all clouds with S/N > 3. The CO cloud parameters are corrected by sensitivity bias and resolution bias, which is originated by the non-zero noise of CO emission and by the instrumental convolutions \citep[finite spatial and spectral resolutions, see][]{Rosolowsky_2006PASP_118_590}. For the calculation of the radius  $R$, CPROPS uses the \cite{Solomon_1987ApJ_319_730S} definition for spherical clouds assuming a factor of $1.91$ to convert the second moments of the emission along the major and minor axes of clouds ($\sigma_{r}$) to $R$. For unresolved clouds along the minor axis, we obtained the radii following \cite{Saldano_2023AA_672A_153S} calculation. The deconvolved $\sigma_{\upsilon}$ is calculated from equation (10) from \cite{Rosolowsky_2006PASP_118_590}, and the FWHM of the spectral line is given by $\sqrt{8\ln{(2)}}\,\sigma_{\upsilon}$.

The CO flux of the clouds is converted to luminosity by mean:

\begin{equation}
\label{eq:luminosity}
\begin{aligned}
 L_{\text{CO}}[\text{K\,km\,s}^{-1}\,\text{pc}^{2}] = \,   
  & F_{\text{CO}}(0\,\text{K})[\text{K\,km\,s}^{-1}\,\text{arcsec}^{2}] 
   (d[\text{pc}])^{2} \\
  &  \times \left(\frac{\pi}{180\,\times\,3600}\right)^{2}
\end{aligned}
\end{equation}

\noindent
where $F_{\text{CO}}(0\text{K})$ is the flux measured at infinite sensitivity, and $d$ is the distance of the galaxy in parsec of $60$ kpc. Another important parameter used to analyze the stability of molecular clouds is the virial mass, given by the formula \citep{Solomon_1987ApJ_319_730S}: 

\begin{equation}
    M_{\text{vir}} = 1040\,\sigma_{\upsilon}^{2}\,R\,\,[M_{\odot}],
    \label{eq:virial_mass}
\end{equation}

\noindent
where $\sigma_{\upsilon}$ and $R$ are the corrected velocity dispersion and radius in parsec, respectively. For this last equation, a self-gravitating sphere with density profile $\rho \propto r^{-1}$ was assumed. The external forces, like magnetic fields and external pressure, are discarded. 

The uncertainties in the properties and derived parameters are estimated by CPROPS with the bootstrapping technique. For the case of the unresolved clouds in which the sizes were recalculated, we estimated the error in the radius and virial mass calculation using the error propagation.

\section{Results} 
\label{sec:results}

\subsection{CO($3-2$) emission in the SMC-Bar}

The distribution of the CO($3-2$) emission in the SMC-Bar is shown in  Fig. \ref{fig:smc_integrated}. The emission is found in many isolated clouds dispersed all over the SMC-Bar. Only two major strong emission concentrations are seen, one located in the SW-bar and the other in the northern N66 region.

In Figs \ref{fig:integrated_SWBar} and \ref{fig:integrated_NEBar}, we show the CO($3-2$) and CO($2-1$) integrated emission maps for the SW-Bar, DarkPk, N66 and NE-Bar (see labels in Fig. \ref{fig:smc_supercam}). The integrated maps have the same spatial resolution of $30''$. For the SW-Bar, the velocity range where the CO emissions were found is $100 - 150$ \kms, and for the DarkPk is between $130 - 165$ \kms. In the northern regions, N66 and NE-Bar, the velocity range of the CO emission is $140 - 170$ \kms\ and $150 - 200$ \kms, respectively.

We found that most of the clouds in the SW-Bar and N66 regions are detected in both $J=3-2$ and $J=2-1$ transitions and only a few clouds show weak or even no CO($3-2$) emission. For example, in N66, the plume-like component to the northeast and the bar structure to the southwest are well seen in the  CO($3-2$) as in the CO($2-1$) map. On the contrary, in the DarkPK, there are only four positions, labeled as A, B, C, and D in the bottom panels of Fig. \ref{fig:integrated_SWBar}, where CO($3-2$) emission is detected. In the NE-Bar region, being the CO($3-2$) map much nosier than in the other regions of the SMC, we also detected four clouds, labeled as A, B, C, and D in the bottom panel of Fig. \ref{fig:integrated_NEBar}.

\subsection{The CO($3-2$)-to-CO($2-1$) ratio in the SMC}
\label{sec:CO_ratio_results}

We determined $R_{32}$ in the SMC using different approaches for each of the regions mapped.  

\begin{figure*}
\centering
\includegraphics[width = 0.48\linewidth]{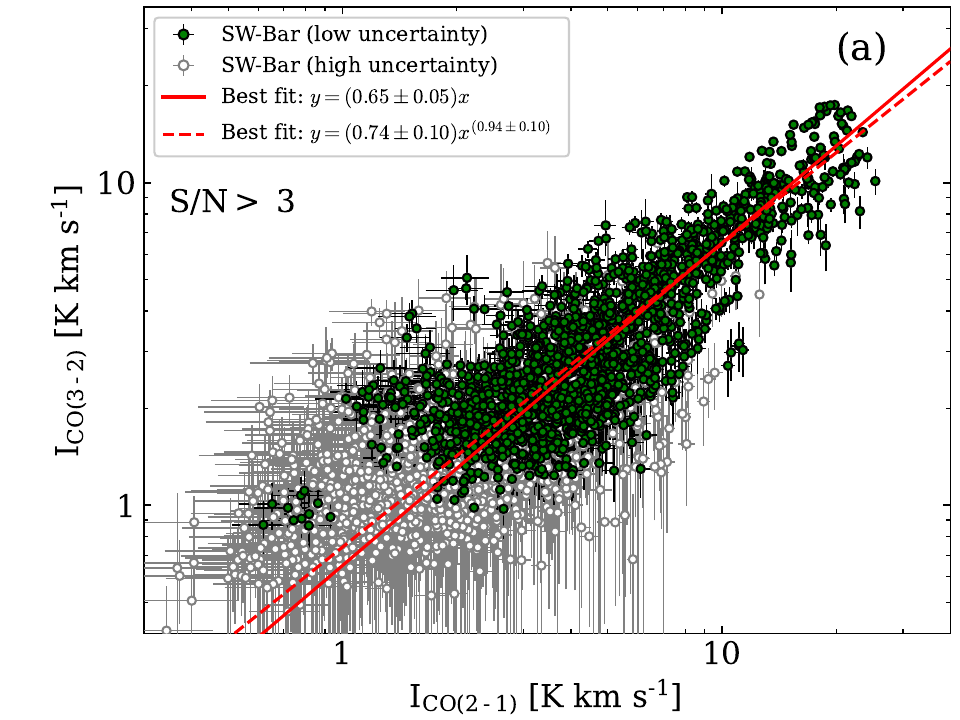}
\includegraphics[width = 0.48\linewidth]{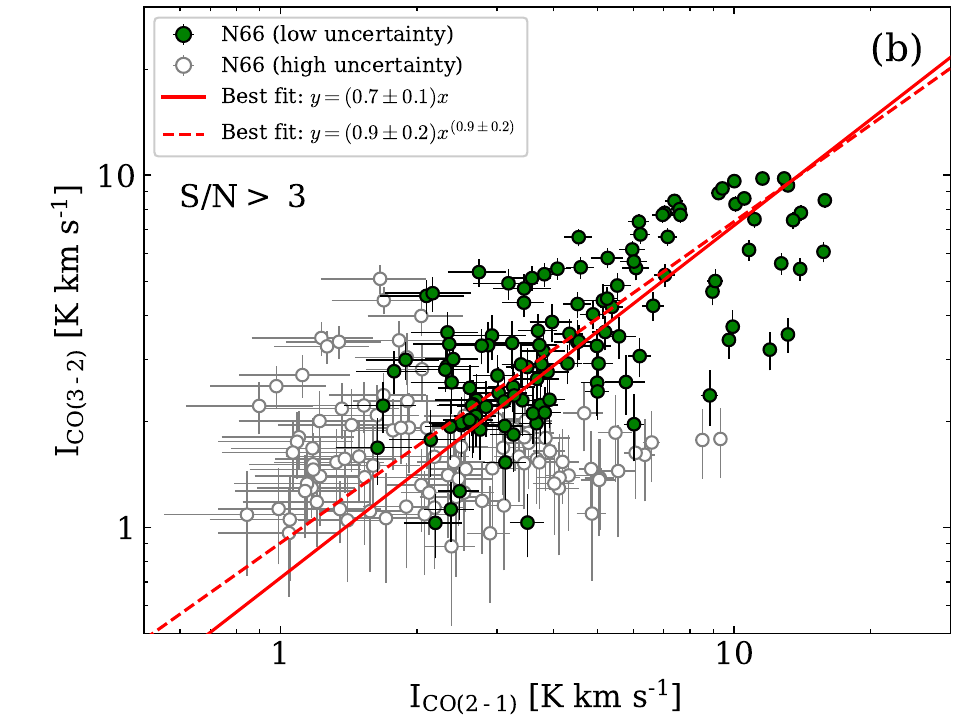}
\caption{Relationship between $I_{\rm CO(3-2)}$ and $I_{\rm CO(2-1)}$ for SW-Bar and N66. The integrated emission in both transitions has S/N $> 3$. The green dots (indicated as low uncertainty) correspond to integrated emission with relative uncertainties $< 20\%$ in both transitions. The white dots indicate integrated emissions with higher relative uncertainties. The solid and dashed red lines show the best fits of the low uncertainty data assuming a linear and a power-law model, respectively.}
\label{fig:Ico_correlation}
\end{figure*}

For the SW-Bar and N66 we used the pixel-by-pixel integrated emissions ($I_{\rm CO}$) of CO($3-2$) and CO($2-1$). In  Fig. \ref{fig:Ico_correlation}, we plotted the correlation of both integrated emissions for each region. This correlation shows a linear trend in the log-log scale for the stronger emission (low uncertainty), i.e., that with lower than $20\%$  uncertainties in both transitions and therefore not noise-dominated. Assuming that $I_{\rm CO(3-2)}$ is proportional to $I_{\rm CO(2-1)}$, and representing this by $y = a\,x$ model, where the slope $a$ gives the integrated CO spectral line ratio ($R_{32}$), we found that the best fit of Fig. \ref{fig:Ico_correlation}a has a slope of $a = 0.65\pm 0.05$ for the SW-Bar, while for N66 in Fig. \ref{fig:Ico_correlation}b, we found a slightly higher slope of $a = 0.7\pm0.1$. If we assumed a power-low model ($y = \alpha\,x^{\,\beta}$), we found that the power-index is $\beta = 0.94\pm0.10$ and $\alpha = 0.74\pm0.10$ for SW-Bar, while for N66 the best fit has a power index of $\beta = 0.9\pm0.2$ and $\alpha = 0.9\pm0.2$. The two model fits are shown in Fig. \ref{fig:Ico_correlation}a and \ref{fig:Ico_correlation}b. The scatter of the two correlations increases for the weaker integrated emission, i.e., for uncertainties higher than $20\%$ (high uncertainty), an effect that could be accounted to the lower S/N ratio of CO($3-2$) compared to that of CO($2-1$). We adopt the linear fit slope for the $R_{32}$ ratio. 

\begin{figure*}
\centering
\includegraphics[width = 0.5\linewidth]{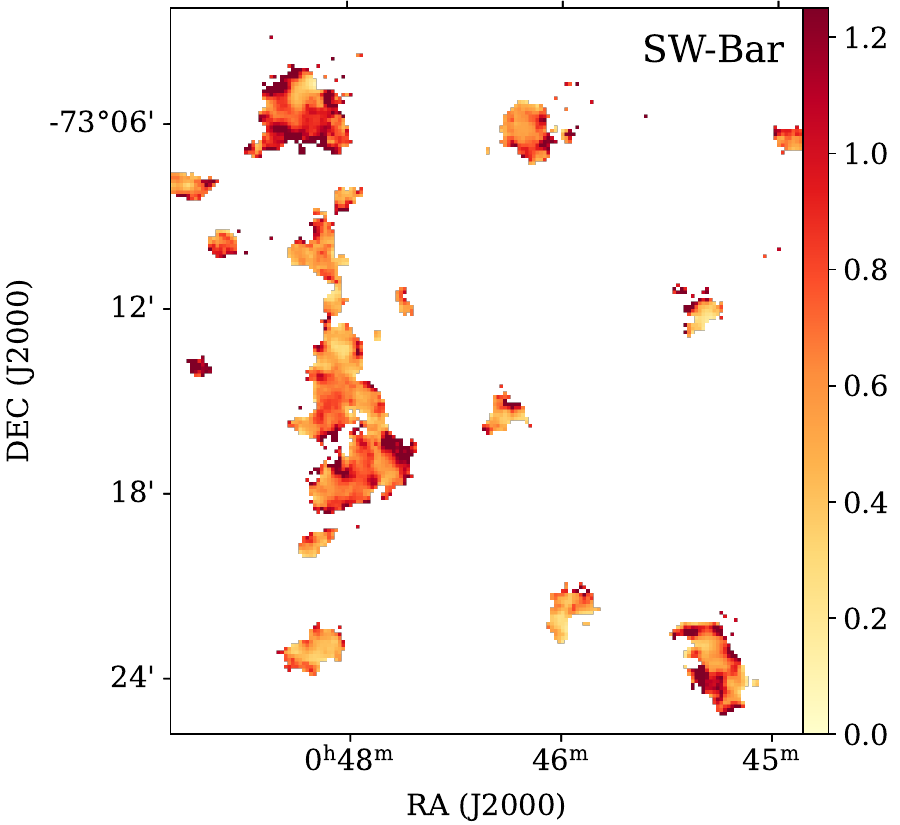}
\includegraphics[width = 0.35\linewidth]{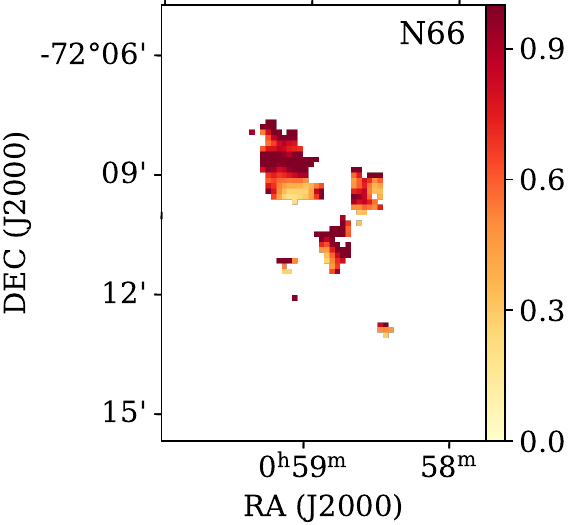}
\caption{$R_{32}$ maps of the SW-Bar and N66 regions at $30''$ resolution. The $R_{32}$ scale is shown on the right of each map. This ratio maps correspond to S/N $> 3$ in integrated intensity in both transitions.}
\label{fig:CO_ratio_maps}
\end{figure*}

\begin{figure*}
\centering
\includegraphics[width = 0.8\linewidth]{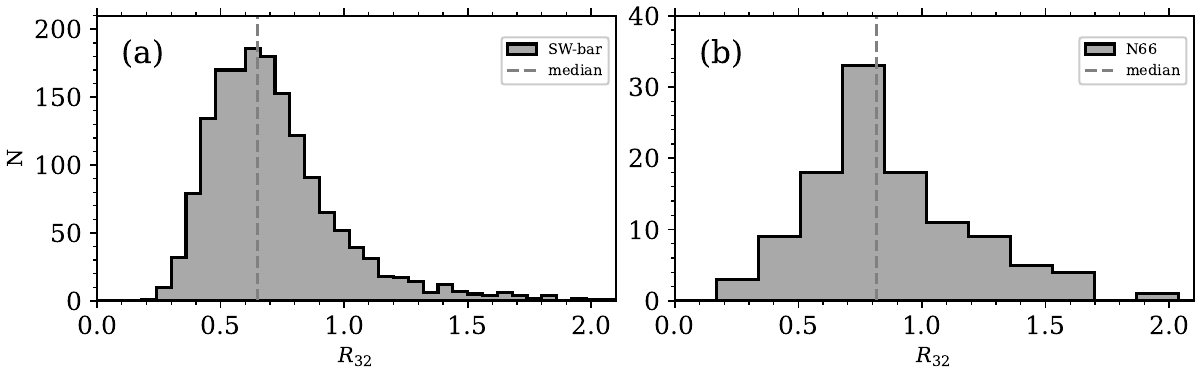}
\caption{Histograms of the pixel-by-pixel $R_{32} =$ [CO($3-2$)/CO($2-1$)] ratio  for SW-Bar and N66. The vertical dashed lines indicate the median $R_{32}$ of $0.65\pm0.30$ and $0.8\pm0.3$ for SW-Bar and N66, respectively. }
\label{fig:CO_ratio_histo}
\end{figure*}

As an alternative method to determine a median ratio $R_{32}$ in SW-Bar and N66, we made histograms from the $R_{32}$ maps (Fig. \ref{fig:CO_ratio_maps}). These histograms are shown in Fig. \ref{fig:CO_ratio_histo} and were made with the pixel values of the $R_{32}$ maps with uncertainties lower than $30\%$. This limit was estimated to be consistent with the integrated CO emission of low uncertainty ($< 20\%$) in both transitions used in Fig. \ref{fig:Ico_correlation}. We found that the median of $R_{32}$ for SW-Bar and N66 is $0.65$ and $0.8$, respectively, both with a standard deviation of $0.3$.

\begin{figure*}
\centering
\includegraphics[width = \linewidth]{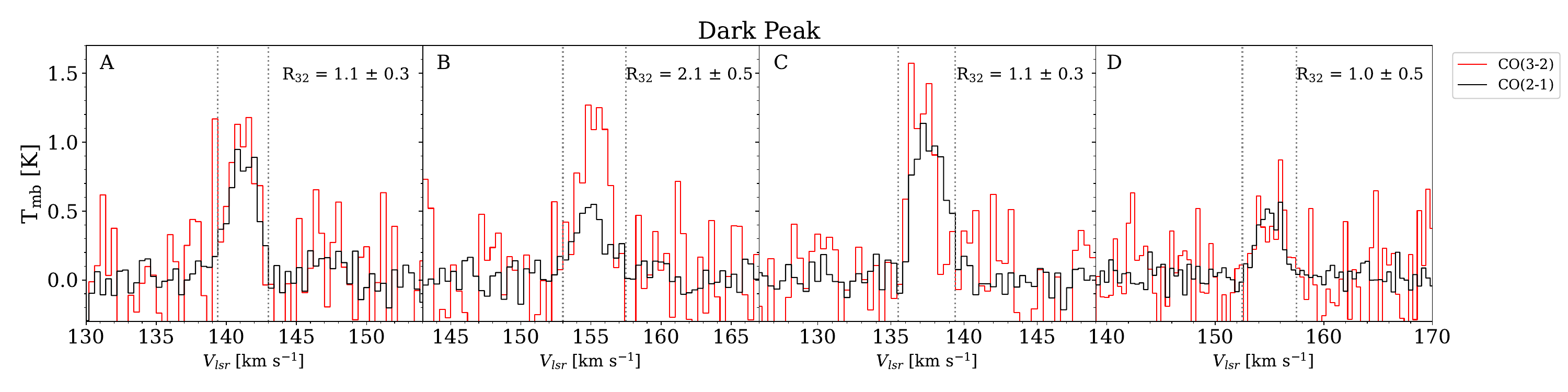}
\includegraphics[width = \linewidth]{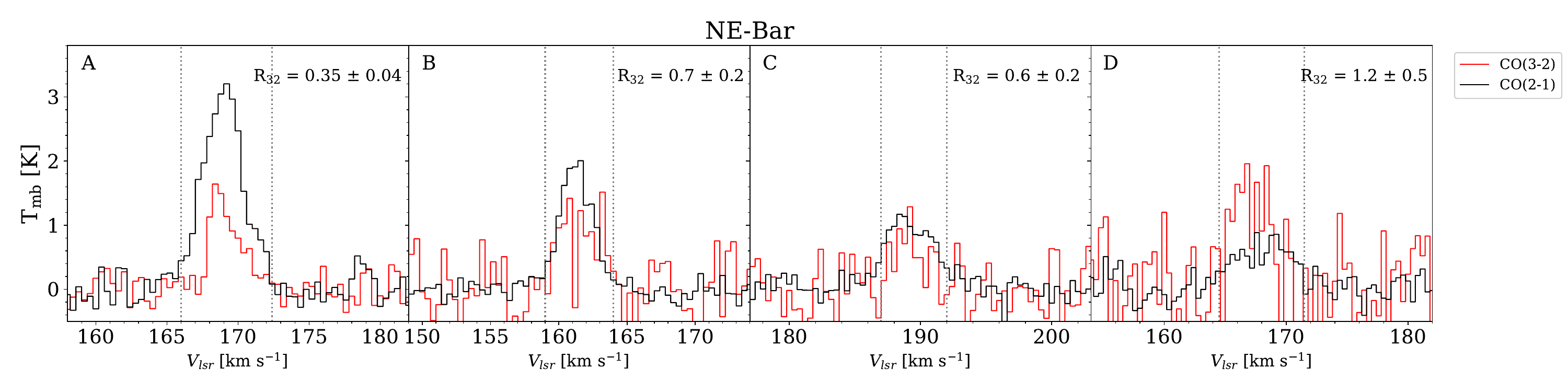}
\caption{CO spectral lines in the transitions $J=3-2$ and $J=2-1$ toward DarkPK (A, B, C, and D) and NE-Bar (A, B, C, and D). The positions where the CO lines were extracted are shown in Figs. \ref{fig:integrated_SWBar} and \ref{fig:integrated_NEBar}. The CO($3-2$) lines are shown in red, and the CO($2-1$) lines in black. The estimated $R_{32}$, using the moment method within the vertical dotted lines, are shown in the right upper corner of each panel.}
\label{fig:CO_lines}
\end{figure*}

For the DarkPK and NE-Bar, we only estimated $R_{32}$ using the spectral lines taken from the peak positions indicated in Figs \ref{fig:integrated_SWBar} and \ref{fig:integrated_NEBar}. We show these spectral lines in Fig \ref{fig:CO_lines}. In the DarkPK, we found that  $R_{32} \simeq 1-2$ in the position A, B, C, and D, with $\sim 20-50\%$ of uncertainties. While in NE-Bar, we found a $R_{32}$ of $0.35\pm0.04$, $0.7\pm0.2$, $0.6\pm0.2$, and $0.9\pm0.4$ for the A, B, C, and D positions, respectively.

\subsection{SMC CO($3-2$) Clouds}

We have identified $225$ CO clouds in the SMC with S/N ratio $> 3$. Out of this total sample, $201$ are well-resolved clouds with $R \ge 1.5$ pc. Increasing the S/N ratio~$\gtrsim$ to ~$5$ then  $20$ clouds are found with only $17$ that have resolved sizes. These $17$ clouds are the brightest, well-resolved CO clouds in the $J = 3-2$ transitions at 6 pc resolution that we found in the SMC. The parameters of these clouds are listed in Table \ref{tab:cprops_parameters_SMC} \footnote{The complete table that includes the parameters of the 225 clouds can be found in the online version.}. 

\begin{table*}
\centering
\caption{Physical parameters of CO($3-2)$ clouds in the SMC}
\begin{threeparttable}
\begin{tabular}{lcc
                S[table-format=1.1]@{\,\( \pm \)\,}
                S[table-format=1.1]
                S[table-format=3.1]@{\,\( \pm \)\,}
                S[table-format=1.1]                
                S[table-format=1.1]@{\,\( \pm \)\,}
                S[table-format=1.1]
                S[table-format=1.1]@{\,\( \pm \)\,}
                S[table-format=1.1]
                S[table-format=2.1]@{\,\(\pm\)\,}
                S[table-format=1.1]
                S[table-format=2.1]@{\,\(\pm\)\,}
                S[table-format=1.1]
                S[table-format=2.1]@{\,\( \pm \)\,}
                S[table-format=2.1]}
\hline
\hline
  ID  &    R.A.    &    Decl.   & 
  \multicolumn{2}{c}{$R$}                  &
  \multicolumn{2}{c}{$V_{\text{lsr}}$}     &
  \multicolumn{2}{c}{$\sigma_{\upsilon}$}  &
  \multicolumn{2}{c}{$T_{\text{peak}}$}    &
  \multicolumn{2}{c}{$I_{\text{CO}(3-2)}$}      &
  \multicolumn{2}{c}{$L_{\text{CO}(3-2)}$}      &
  \multicolumn{2}{c}{$M_{\text{vir}}$}     \\
      & (hh:mm:ss.s) & (dd:mm:ss) & 
      \multicolumn{2}{c}{(pc)}      &
      \multicolumn{2}{c}{(\kms)}    &
      \multicolumn{2}{c}{(\kms)}    & 
      \multicolumn{2}{c}{(K)}       &
      \multicolumn{2}{c}{(caption)} &
      \multicolumn{2}{c}{(caption)} &
      \multicolumn{2}{c}{(10$^{3}$\Msun)} \\
\hline
 14 & 00:46:27.38 & -73:29:49.6 &  \multicolumn{2}{c}{...} &  82.5 & 1.3 & 1.1 & 0.5 & 2.0 &  0.4 &   0.4 & 0.1 &   2.3 & 0.5 &   \multicolumn{2}{c}{...} \\
 17 & 00:46:40.33 & -73:06:10.5 &  7.3 & 1.1 & 126.0 & 1.2 & 1.0 & 0.1 & 2.5 &  0.3 &   3.5 & 0.4 &  19.1 & 2.4 &   8.2 &  2.6 \\
 28 & 00:47:43.51 & -73:17:06.5 &  4.5 & 1.3 & 121.1 & 1.8 & 1.6 & 0.2 & 1.4 &  0.2 &   1.9 & 0.3 &  10.6 & 1.5 &  11.5 &  5.0 \\
 31 & 00:47:54.46 & -73:17:20.0 &  8.5 & 0.8 & 120.8 & 1.7 & 1.5 & 0.2 & 1.7 &  0.2 &   4.2 & 0.3 &  23.2 & 1.8 &  18.8 &  5.1 \\
 39 & 00:48:05.71 & -73:17:51.1 &  5.6 & 1.2 & 119.3 & 2.3 & 2.0 & 0.3 & 1.5 &  0.2 &   2.9 & 0.3 &  16.0 & 1.7 &  22.7 &  7.7 \\
 42 & 00:48:08.18 & -73:14:47.3 &  8.8 & 1.0 & 120.5 & 2.4 & 2.1 & 0.3 & 1.5 &  0.2 &   3.9 & 0.4 &  21.3 & 2.0 &  39.3 & 11.9 \\
 56 & 00:48:17.75 & -73:05:17.2 &  3.8 & 1.4 & 110.8 & 1.1 & 0.9 & 0.5 & 1.0 &  0.2 &   0.4 & 0.1 &   2.2 & 0.7 &   3.4 &  4.0 \\
 58 & 00:48:20.65 & -73:05:50.3 & 13.0 & 0.8 & 113.6 & 3.3 & 2.8 & 0.2 & 1.6 &  0.2 &  11.4 & 0.5 &  62.8 & 2.6 & 108.3 & 18.0 \\
 59 & 00:48:22.06 & -73:10:13.8 &  5.2 & 1.9 & 117.6 & 1.2 & 1.0 & 0.3 & 1.2 &  0.2 &   0.9 & 0.2 &   4.9 & 0.9 &   5.2 &  3.6 \\
 67 & 00:48:41.98 & -73:00:49.3 &  2.8 & 1.6 & 126.9 & 1.0 & 0.8 & 0.4 & 2.1 &  0.4 &   0.3 & 0.1 &   1.7 & 0.6 &   2.0 &  2.1 \\
 69 & 00:48:56.39 & -73:09:54.1 &  3.1 & 1.7 & 124.7 & 1.0 & 0.8 & 0.2 & 1.3 &  0.2 &   0.6 & 0.1 &   3.5 & 0.6 &   2.2 &  1.7 \\
 73 & 00:49:04.87 & -72:47:39.3 &  3.9 & 1.4 & 133.6 & 1.4 & 1.1 & 0.5 & 2.1 &  0.4 &   0.3 & 0.1 &   1.5 & 0.6 &   5.3 &  5.3 \\
 85 & 00:49:44.72 & -73:24:27.3 &  4.2 & 1.4 & 144.1 & 1.1 & 1.0 & 0.4 & 1.5 &  0.3 &   0.8 & 0.2 &   4.3 & 1.1 &   4.0 &  3.4 \\
107 & 00:52:22.74 & -72:22:07.0 &  \multicolumn{2}{c}{...} & 194.5 & 1.9 & 1.6 & 0.5 & 5.5 &  1.1 &   0.9 & 0.1 &   4.7 & 0.8 &   \multicolumn{2}{c}{...} \\
145 & 00:58:35.61 & -72:27:31.1 &  3.7 & 1.4 & 122.1 & 1.7 & 1.4 & 0.3 & 3.1 &  0.5 &   1.1 & 0.2 &   6.3 & 1.0 &   7.9 &  4.5 \\
163 & 00:59:19.65 & -72:08:42.6 &  7.9 & 1.2 & 159.3 & 1.8 & 1.5 & 0.2 & 3.2 &  0.6 &   2.9 & 0.4 &  16.0 & 2.4 &  18.6 &  5.8 \\
165 & 00:59:21.01 & -72:08:54.4 &  6.6 & 1.4 & 163.2 & 1.6 & 1.4 & 0.3 & 2.9 &  0.6 &   1.5 & 0.2 &   8.2 & 1.3 &  12.7 &  6.4 \\
185 & 01:01:15.16 & -72:45:18.7 &  3.0 & 1.6 & 162.6 & 1.2 & 1.0 & 0.3 & 6.1 &  1.3 &   1.1 & 0.2 &   6.0 & 0.9 &   3.4 &  2.7 \\
195 & 01:01:46.64 & -72:14:34.1 &  \multicolumn{2}{c}{...} & 156.4 & 0.9 & 0.7 & 0.3 & 4.6 &  0.9 &   0.5 & 0.2 &   2.7 & 1.0 &   \multicolumn{2}{c}{...} \\
201 & 01:03:06.89 & -72:03:46.9 &  3.4 & 1.8 & 168.9 & 1.3 & 1.1 & 0.5 & 3.8 &  0.8 &   1.0 & 0.3 &   5.4 & 1.6 &   4.3 &  4.2 \\
\hline
\hline
\end{tabular}
\begin{tablenotes}
   \item {\bf Note:} $I_{\text{CO(3-2)}}$ and $L_{\text{CO(3-2)}}$ are in unit of $10^{2}$\,K\,\kms\ and $10^{2}$\,K\,\kms\,pc$^{2}$, respectively.\\
   This table only lists those clouds with S/N $\gtrsim 5$. A complete version of the table, with the $225$ clouds identified with S/N $> 3$, can be found in the online version.  The ID numbers in col. 1 follow the order of the complete table.
\end{tablenotes}
\end{threeparttable}
\label{tab:cprops_parameters_SMC}
\end{table*}

\begin{figure*}
\centering
\includegraphics[width = 0.33\linewidth]{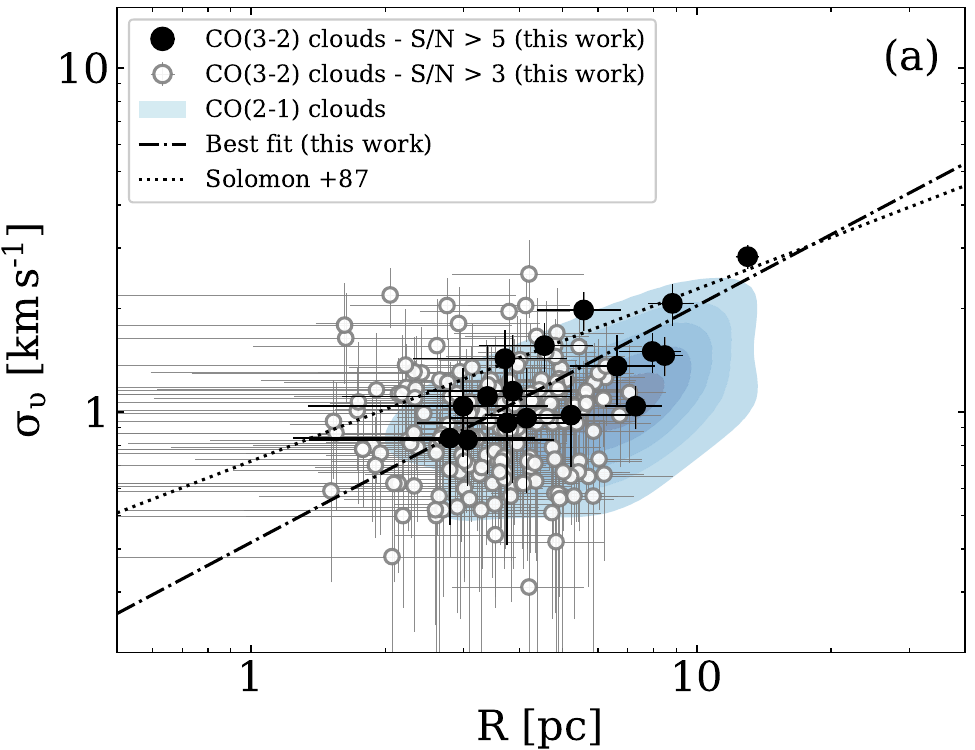}
\includegraphics[width = 0.33\linewidth]{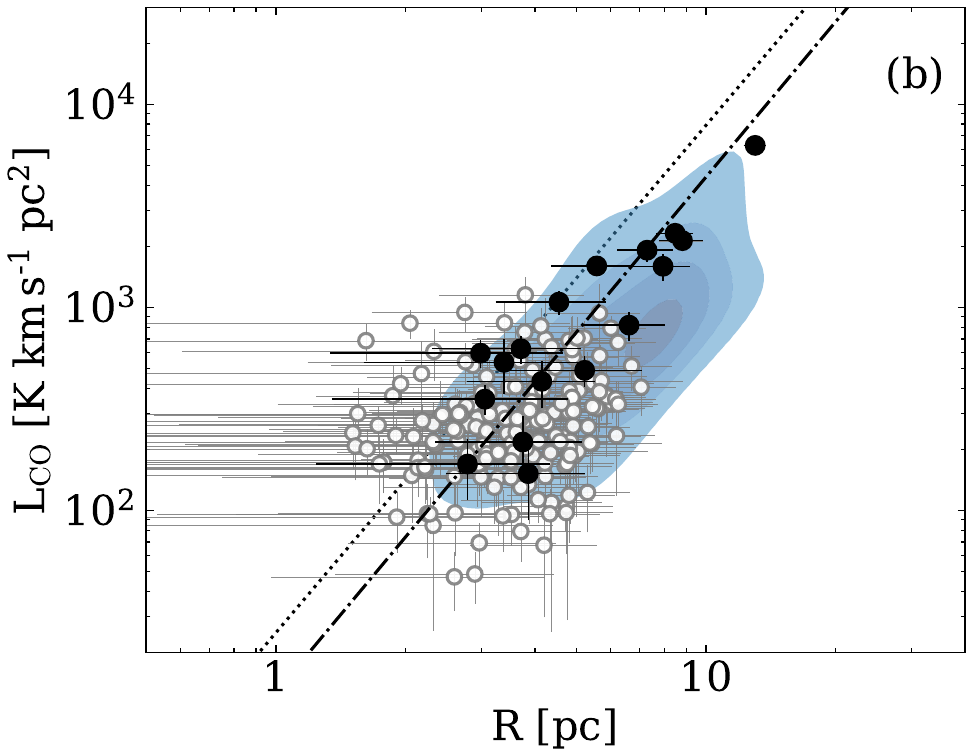}
\includegraphics[width = 0.33\linewidth]{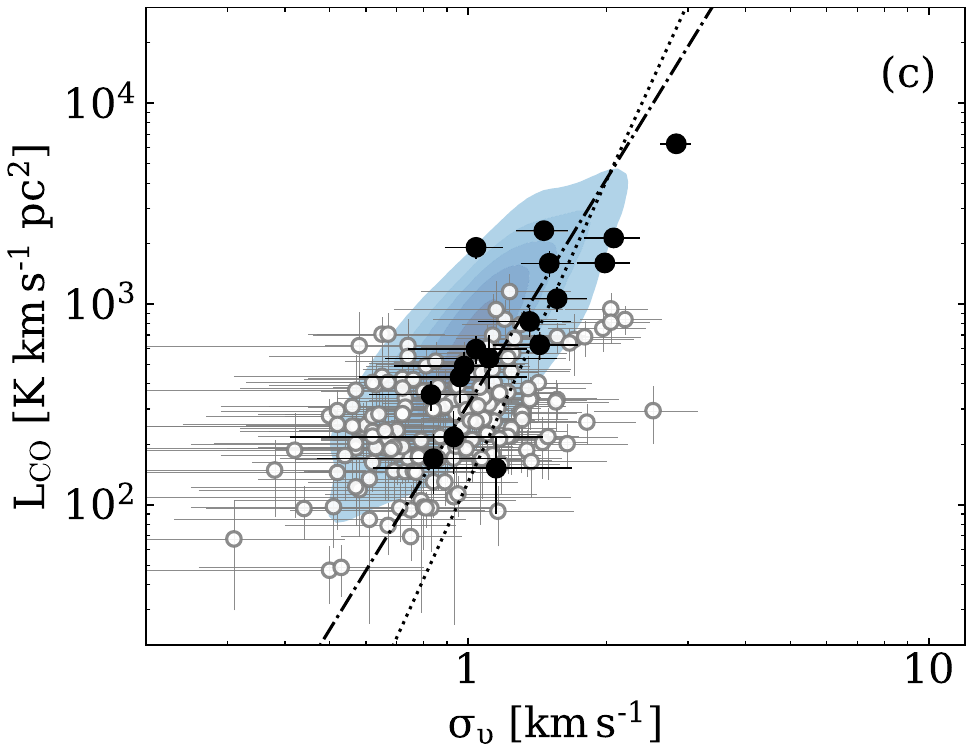}
\includegraphics[width = 0.33\linewidth]{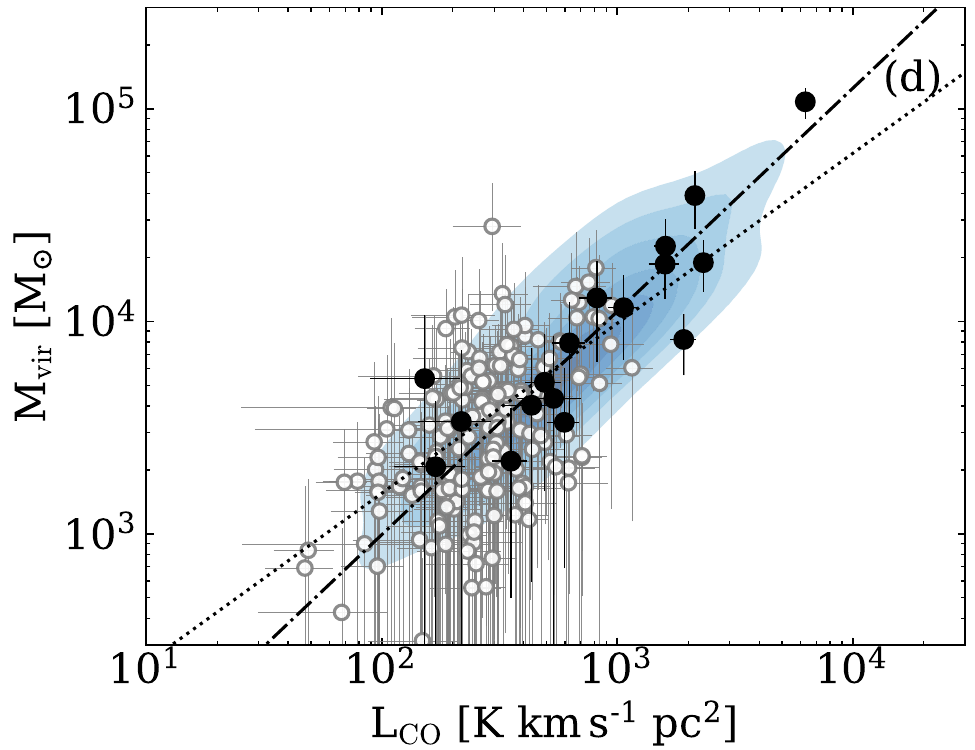}
\includegraphics[width = 0.33\linewidth]{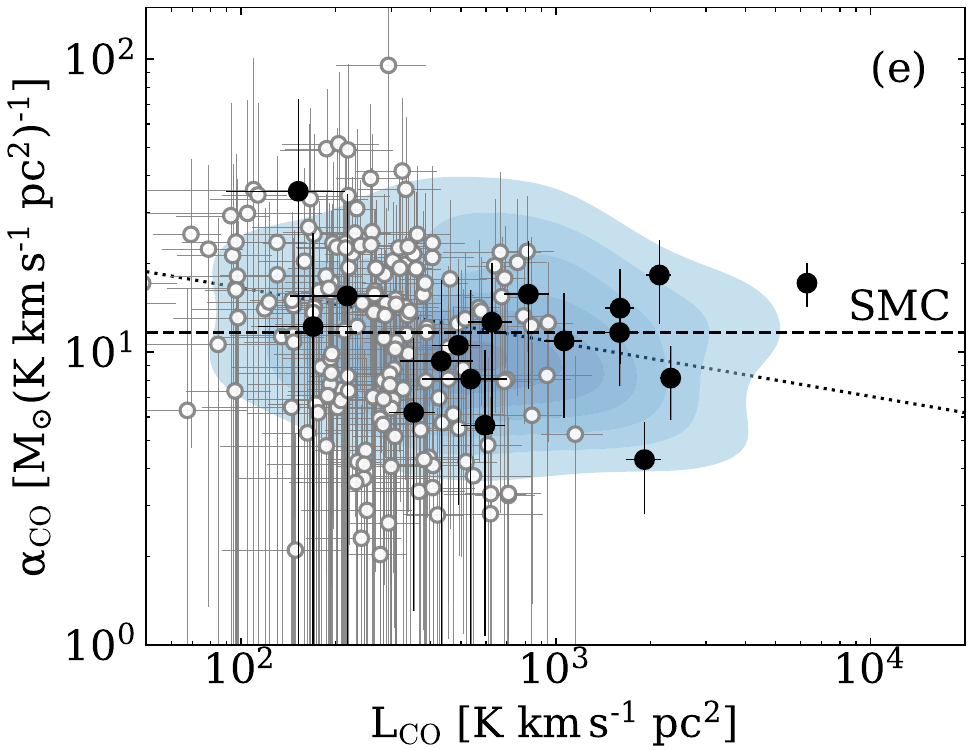}
\caption{Scaling relations of the CO($3-2$) clouds parameters in the SMC: panel a) the size-linewidth relation ($\sigma_{\upsilon}-R$), panel b) the luminosity-size relation ($R-L_{\rm CO}$), panel c) the luminosity-linewidth relation ($\sigma_{\upsilon}-L_{\rm CO}$), panel d) the luminosity-virial mass relation ($L_{\rm CO}-M_{\rm vir}$), and panel e) the CO-to-H$_2$ conversion factor-luminosity relation ($\alpha_{\rm CO} - L_{\rm CO}$). The black dots correspond to clouds with S/N ratio $\gtrsim 5$, while the gray dots correspond to clouds with S/N between $3$ and $5$.  For comparison, we plot the distribution of the CO($2-1$) clouds in a blue color palette scale showing the concentration of the CO($2-1$) clouds \citep{Saldano_2023AA_672A_153S}. In all the panels, the best fits for the brightest CO($3-2$) clouds are indicated by dot-dashed lines. The \cite{Solomon_1987ApJ_319_730S} relation is indicated by dotted lines. In panel (e), the dashed horizontal line indicates the median  $\alpha_{\rm CO(3-2)}$ of $12.6_{-7}^{+20}$ \Xco.}
\label{fig:SMC_scaling_relation}
\end{figure*}

We used all the CO($3-2$) clouds identified by CPROPS to show the scaling relations between their properties, i.e., size-linewidth, Luminosity-size, luminosity-velocity dispersion, and virial mass-luminosity. In Figure \ref{fig:SMC_scaling_relation}, we plotted the scaling relations for the CO($3-2$) clouds with S/N ratio $\gtrsim 5$ in black dots and the CO($3-2$) clouds with S/N ratio between $3$ and $5$ in white dots. However, we only performed linear fits in the log-log scales to these scaling relations using the brightest, well-resolved clouds (S/N $\gtrsim 5$), as the parameters of clouds with low S/N ratios tend to be more uncertain \citep{Rosolowsky_2006PASP_118_590} and will highly affect the linear fit in the scaling relation \citep[see][]{Wong_2011ApJS_197s}. We followed the procedure indicated in \cite{Saldano_2023AA_672A_153S} and found the following relationships:

\begin{align}
    &\log{\sigma_{\upsilon}} ~~=~ (-0.38\pm0.09) + (0.7_{-0.1}^{+0.1})\log{R} \\
    &\log{L_{\rm CO}} =~ (1.1\pm0.3) + (2.5_{-0.4}^{+0.3})\log{R}\\
    &\log{L_{\rm CO}} =~ (2.5\pm0.1) + (3.7_{-0.8}^{+1.0})\log{\sigma_{\upsilon}}\\
    &\log{M_{\rm vir}} =~ (0.9\pm0.5) + (1.0_{-0.2}^{+0.1})\log{L_{\rm CO}}
\end{align}

In each of the five panels of Figure \ref{fig:SMC_scaling_relation}, we included the best fit obtained. For comparison, we plotted in blue the distribution of the CO($2-1$) clouds which contains more than $90\%$ of the CO($2-1$) clouds with S/N $> 5$ \citep{Saldano_2023AA_672A_153S}. This dot density map is plotted in steps of 0.8, 0.4, 0.5, and 0.3 dex per unit area in panels (a) to (d), respectively. We included in the panels the inner Milky Way relationship (Solomon et al 1986).  

The size-linewidth relation shows that the CO($3-2$) clouds are below the Milky Way clouds (solid black line) of similar size by a factor of $1.3\pm0.3$. The comparison of the luminosity scaling relations between the SMC CO($3-2$) and the Milky Way shows that the SMC clouds are under-luminous for similar sizes by a factor of $1.9\pm1.0$ and over-luminous for similar velocity dispersion by a factor of $1.7\pm1.6$. The luminosity-virial mass relation (panel d) shows that for luminosities larger than  $L_{\rm CO} \sim 3-4\times10 ^3$, the CO($3-2$) tend to have larger virial masses than their Milky-Way counterpart at similar luminosity, while for clouds less luminous the virial masses show a larger dispersion in their value.
Thus, the scaling relations of the CO($3-2$) are similar (within the error) to the CO($2-1$) clouds scaling relations found by \cite{Saldano_2023AA_672A_153S}.

Assuming virial conditions for the CO($3-2$) clouds, we can estimate the CO-to-H$_2$ conversion factor ($\alpha_{\rm CO(3-2)}$) as the ratio between the virial mass and CO luminosity. In panel (e) of Fig. \ref{fig:SMC_scaling_relation}, we plot the conversion factor as a function of the luminosity. We did not find an evident correlation between the CO luminosity and $\alpha_{\rm CO(3-2)}$ (Spearman's $\rho = -0.02$). The distribution of the conversion factor of the CO($3-2$) bright clouds has a median value of $11.7_{-4}^{+5}$ \Xco. Including those clouds with S/N ratio between $3$ and $5$, then the median conversion factor is $12.6^{+10}_{-7}$ \Xco.

\section{Discussion}
\label{sec:discussion}

\subsection{The $R_{32}$ in the SMC}

We found a median $R_{32}$ of $0.65$ and an average $R_{32}$ of $0.7$ for the SMC-Bar, with a standard deviation of $0.3$. These values are similar to $R_{32} \simeq 0.6-0.8$, found in low-metallicity galaxies with metallicities ranging between $0.1-0.5~Z_{\odot}$ \citep{Hunt_2017_AA_606A_99H} and somehow higher than the mean ratio $R_{32} \simeq 0.5$ found in the local group of low-mass dwarf galaxies \citep[]{Leroy_2022ApJ_927_149L}. 

Depending on the environment of the observed region in the SMC, we found variations of the $R_{32}$ values. For example, we estimated a median value of $0.65$ in the SW-Bar which is similar to the value in the quiescent SMCB1\#1 cloud located in the south-west part of the SMC obtained by \cite{Bolatto_2005ApJ_633_210B}, although measured at a coarser $109''$ ($32$ pc) resolution than ours at $30''$ ($9$ pc). However, we found that more evolved and active star formation regions \citep[N22, SWBarN, SWBarS, associated to multiple HII regions, see][]{Jameson_2018_ApJ_853_111J} have median $R_{32} = 0.7 - 0.9$, higher than the quiescent cloud (see Fig. \ref{fig:SW_ratio_Halpha}). This is also the case for N66, the brightest HII region of the SMC, which shows a  median $R_{32} = 0.8$ value.  
 
Inspecting the distribution of the $R_{32}$ values in the clouds, we found a high dispersion of $R_{32}$ in SW-Bar and in N66, between $0.2$ and $2.0$ (see Fig. \ref{fig:CO_ratio_histo}) which may be an indication of a dependence of $R_{32}$ on environmental local conditions \citep{Penialoza_2018_MNRAS_475_1508P}. Figs. \ref{fig:SW_ratio_Halpha} and \ref{fig:NE_ratio_Halpha} show that the highest values of $R_{32}$ are associated with strong emission at $8~\mu$m and H$\alpha$, while lower values than $0.65$ are found in regions that are either in the line-of-sight of the weakest HII regions or not associated to any HII region. \cite{Penialoza_2018_MNRAS_475_1508P} modeled the ISM evolution affected by the interstellar radiation field (ISRF) and found changes in the CO line ratios depending on the environmental conditions. They showed that $R_{32}$ can increase by a factor up to $\sim 1.5-2.0$ as the ISRF increases two orders of magnitudes, which would be consistent with our findings.

In the DarkPK, the four points (A, B, C, and D) where we estimated $R_{32}$ (see Figs. \ref{fig:integrated_NEBar} and \ref{fig:CO_lines}) give $R_{32} = 1-2$. Such high ratios were not expected, as the DarkPK has no signs of active star formation. This region is dominated by the ISRF and has a low range of visual extinction, $A_V < 1$ \citep{Jameson_2018_ApJ_853_111J}. In this region, the far-ultraviolet photons from evolved stellar sources controlling the gas heating and chemistry would be dissociating more rapidly the CO($2-1$) in the external part of the H$_2$ envelopes than the CO($3-2$) located in the innermost part of the molecular cloud. In this scenario, there would be a CO($2-1$) deficiency increasing $R_{32}$ above the median value. In the NE-Bar region, the integrated CO spectral line ratio towards cloud A is $R_{32} = 0.35\pm0.04$, while $R_{32}$ towards cloud B, C, and D is $R_{32} \ge 0.6$. The cloud B is on the low $R_{32}$ range found and it is not associated neither with strong $8~\mu$m emission nor HII regions, meanwhile, clouds B, C, and D, are on the high $R_{32}$ range, and are associated to the strongest peak emission at $8~\mu$m and H$\alpha$ (see Fig. \ref{fig:NE_ratio_Halpha}).

\begin{figure*}[!ht]
    \centering
    \includegraphics[width=0.37\textwidth]{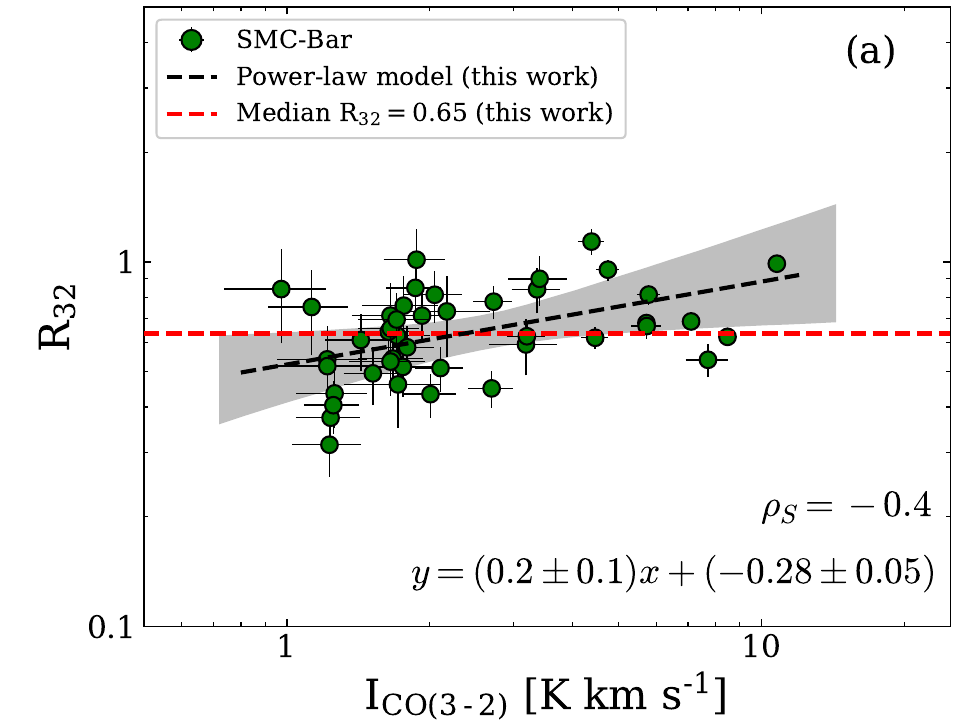}
    \includegraphics[width=0.37\textwidth]{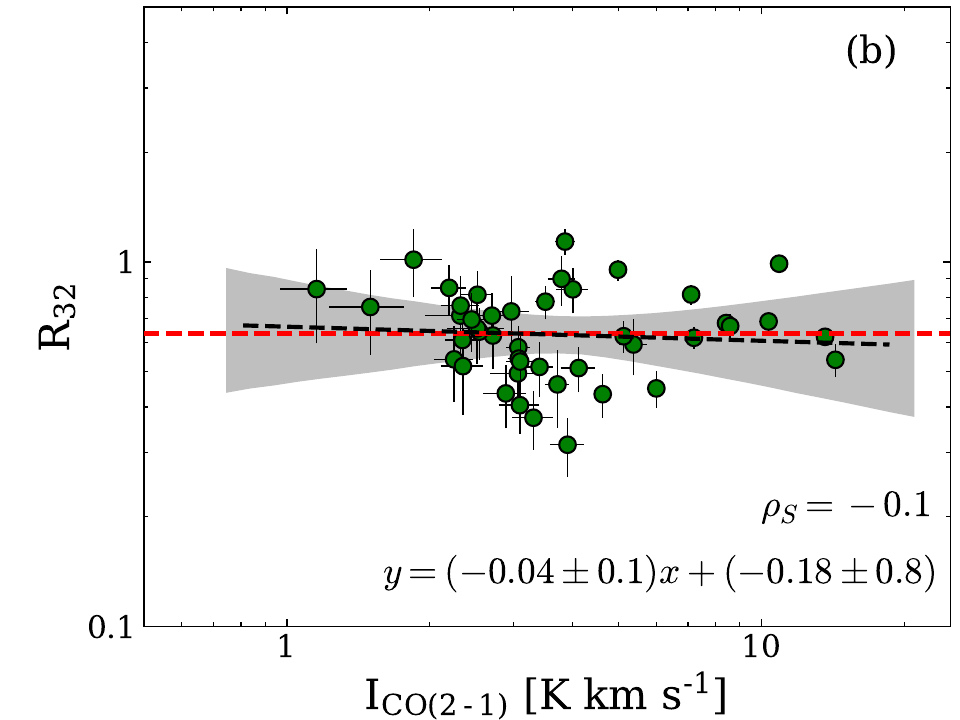}
    \includegraphics[width=0.37\textwidth]{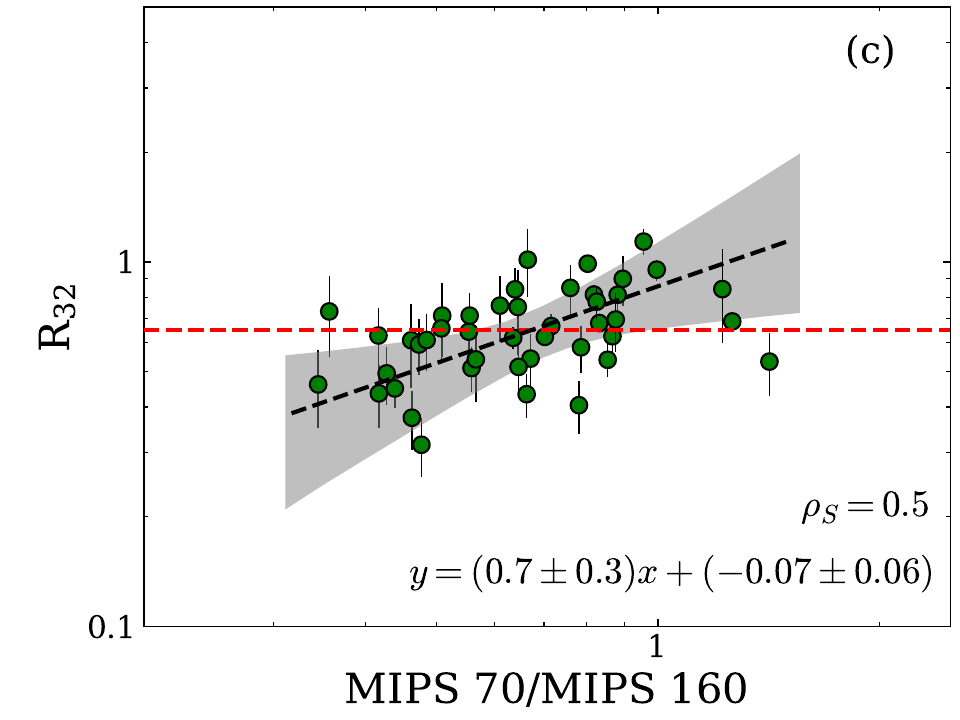}
    \includegraphics[width=0.37\textwidth]{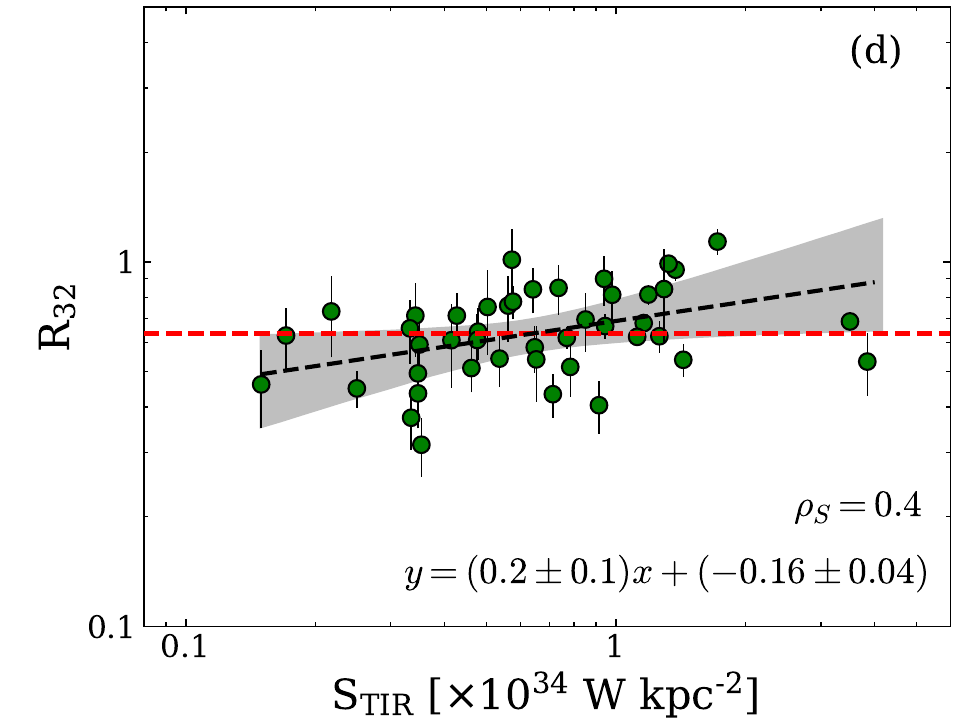}
    \caption{Environment dependence of $R_{32}$. The ratio (with uncertainty lower than $\sim 25\%$) is correlated with the local intensity of CO($3-2$) and CO($2-1$) in panels (a) and (b), respectively. In panel (c), $R_{32}$ as a function of the far-infrared color of the Spitzer MIPS bands \citep{Gordon_2011_AJ_142_102G}, and in panel (d), $R_{32}$ as a function of the TIR surface brightness from Spitzer and Herschel bands \citep{Gordon_2011_AJ_142_102G,Gordon_2014_ApJ_797_85G}. The red and black dashed lines in all panels indicate the median value of $R_{23}$ and the best fit, respectively. The Spearman's correlations ($\rho_{S}$) and the best-fit equations (in log-log scale) are located in the right bottom corners.}
    \label{fig:R32_vs_environment}
\end{figure*}

\subsection{Environment dependence of $R_{32}$ in the SMC}

In order to explore the dependence of $R_{32}$ with observational properties, we made correlations with CO emission, far IR (FIR) color, and total IR (TIR) surface brightness which trace star formation activity. We used the intensity of CO($2-1$) and CO($3-2$) shown in our study, the Herschel bands ($100-250~\mu$m) from HERITAGE \citep{Gordon_2014_ApJ_797_85G}, and the Spitzer MIPS $24$, $70$ and $160~\mu$m from SMC-SAGE \citep{Gordon_2011_AJ_142_102G}. We convolved the images to the HERITAGE Herschel maps to have a common resolution and spatial grid maps. In this analysis, the parameters estimated by the integrated maps of both SW-bar and N66 CO data cubes were used jointly and are referred to as SMC-Bar in the plots. 

We show the correlation of $R_{32}$ with the local integrated intensity of  CO($3-2$) and CO($2-1$) in the panels (a) and (b) of Fig. \ref{fig:R32_vs_environment}, respectively. We found a very weak correlation of $R_{32}$ with CO($3-2$), with a  Spearman's correlation $\rho_{\rm S} = 0.4$.  At high brightness temperature in the  CO $J=3-2$ transition, reliable values of $R_{32}$ (lower uncertainty than $\sim 25\%$) tend to have values higher than the median value. However, in low brightness temperature regions, $R_{32}$ shows a higher dispersion giving a hint of values lower than the median. On the contrary, $R_{32}$ do not correlate with CO($2-1$) (Spearman's correlation is $\rho_{\rm S} = -0.1$).

The correlation of $R_{32}$ with the FIR color $[70/160]$ of the Spitzer bands is shown in panel (c) of Fig. \ref{fig:R32_vs_environment}, with a Spearman's coefficient of $\rho_{\rm S} = 0.5$. The $R_{32}$ increases for higher values of the FIR color. For environments with FIR colors $[70/160]$ larger than $0.8$, most of the $R_{32}$ have values above the median of $0.65$ determined for the SMC-Bar. This correlation is similar to finding in nearby disk galaxies \citep{denBrok_2021_MNRAS_504_3221D}, based on the integrated spectral line CO($2-1$) to CO($1-0$) ratio ($R_{21}$) instead. As the FIR color trace dust temperature and also interstellar radiation field (ISRF) strength, $R_{32}$ in the SMC-Bar is found higher in active star-forming regions, where the ISRF, dust temperature, and the density of the gas should be higher.

Finally, we correlated $R_{32}$ with the TIR surface brightness in the SMC-Bar as shown in panel (d) of Fig. \ref{fig:R32_vs_environment}. To determine the TIR surface brightness we have followed the same method described in  \cite{Jameson_2018_ApJ_853_111J}, where the Spitzer MIPS $24$ and $70~\mu$m bands are combined to the Herschel $100$, $160$, and $250~\mu$m bands by using the equation $S_{\rm TIR} = \sum\, c_{i}\,S_{i}$. $S_{\rm TIR}$ is the TIR surface brightness, and $S_{i}$ is the brightness in the given Spitzer and Herschel bands $i$. Both are in the unit of W\,kpc$^{-2}$ and $c_{i}$ is the calibration coefficient from combined brightness provided by \cite{Galametz_2013_MNRAS_431_1956G}. We found a moderate correlation between $R_{32}$ and the TIR surface brightness (Spearman's coefficient $\rho_{\rm S} \simeq 0.4$). $R_{32}$ tends to be lower than the median value for low TIR surface brightness, while $R_{32}$ tends to increase above the median value for higher TIR surface brightness. The TIR surface brightness scales with the molecular gas surface density, where we expect that the star formation activity may be embedded. Our finding indicates that $R_{32}$ in the SMC-Bar tends to be higher in very active, and dense star-forming regions.

These correlations could be useful to consider when using $R_{32}$ as a diagnostic for environmental properties in external galaxies. To improve these correlations, higher sensitivity observations are needed as these will allow us to increase the range in the observed properties. Moreover, the coarser resolution of the Spitzer and Herschel data used does not allow us to study the smaller structures resolved in this work since the ISM properties are averaged over larger areas.

Considering these caveats, our results indicate that values of $R_{32}$ typically $\sim 0.7-1.0$, in general, were found toward active star-forming regions. These regions are usually excited by HII regions and/or shock emission. While lower $R_{32}$ ratios ($\lesssim 0.6$) were found in more quiescent regions. This behavior is consistent with those found in other galaxies and in semi-analytic studies of the ISM, in which $R_{32}$ (including ratios in other transitions) increase in more active star-forming regions and/or denser and hotter gas, poorly shielded by dust \citep{Penialoza_2018_MNRAS_475_1508P,Celis_2019_AA_628A_96C,denBrok_2021_MNRAS_504_3221D,Leroy_2022ApJ_927_149L}.

\subsection{Scaling Relation in the SMC}

We identified $225$ CO clouds in the $J=3-2$ transition at 6 pc resolution, and we determined their main properties, such as $R$, $\sigma_{\upsilon}$, $L_{\rm CO}$, $M_{\rm vir}$. However, we analyzed the scaling relationships for $17$ high S/N ($\gtrsim 5$) ratio clouds due to the limited sensitivity of our CO($3-2$) survey. Despite the low number of clouds at high S/N ratios, our results are consistent with the study of the scaling relations in the SMC obtained at different spatial resolutions, from 1 to 100 pc \citep[][]{Bolatto_2008ApJ_686,Saldanio_2018_BAAA_60_192S,Saldano_2023AA_672A_153S,Kalari_2020MNRAS_499_2534K,Ohno_2023arXiv230400976O}. It seems that independent of the $J$-transitions and the cloud identification algorithm used, the cloud properties in the scaling relationships would be intrinsic to the SMC.

For example, in the size-linewidth relation (Fig. \ref{fig:SMC_scaling_relation}a), we found that the CO($3-2$) clouds are below the Milky-Way clouds of similar size by a factor of $1.3$. In the CO($2-1$) survey of the SMC,  \cite{Saldano_2023AA_672A_153S} indicate that such departure is by a factor of $\sim 2$, and they explained that it might be due to the CO cloudlets in the SMC, which can not trace the total turbulence of larger H$_{2}$ envelopes \citep[see][]{Bolatto_2008ApJ_686}. They also discard a deficit of turbulent kinetic energy in CO clouds to explain such departure in the size-linewidth relation since these clouds would be gravitationally bounded. \cite{Ohno_2023arXiv230400976O} showed similar results in their CO($2-1$) survey at $2$ pc resolution in the SMC, finding a departure factor of $1.5$ below the Milky-Way trend. These authors ascribe this departure to lower column densities (by a factor of $2$) than those in the Milky Way clouds of similar size. 

Finally, we found that $\alpha_{{\rm CO}(3-2)}$ of our SMC clouds showed a flattened trend with $L_{\rm CO(3-2)}$ (Fig. \ref{fig:SMC_scaling_relation}e) between $10^{2}-10^{4}$ \Lco. We determined a median value of $11.7_{-4}^{+5}$ \Xco\ at $6$ pc resolution for the brightest clouds (S/N $\gtrsim 5$). Similar value was found considering the $225$ CO($3-2$) clouds with S/N $> 3$, giving a $\alpha_{{\rm CO}(3-2)}$ of $12.6_{-7}^{+10}$ \Xco. The  $\alpha_{{\rm CO}(3-2)}$ values obtained are not  very different to the virial-based $\alpha_{{\rm CO}(2-1)}$ of $10.5\pm5$ \Xco\ estimated at $9$ pc resolution in the CO($2-1$) survey of the SMC by \cite{Saldano_2023AA_672A_153S}.  Convolving the SW and N66  CO($3-2$) maps to 9 pc and  comparing  with the CO($2-1$) in the same regions, we found that $\alpha_{{\rm CO}(3-2)}$ at $9$ pc resolution is $14.5_{-7}^{+8}$ \Xco, a bit higher than that estimated at $6$ pc. Despite the uncertainties in the determination of conversion factors, $\alpha_{{\rm CO}(3-2)}$ tend to be higher than $\alpha_{{\rm CO}(2-1)}$ by a factor of $\sim 1.4$, which agrees with our estimation of $1/R_{32} \sim 1.5$. This result is expected for CO clouds that tend to have lower luminosities at higher J-transitions, while the virial mass would not change considerably.
 
\section{Conclusions}
\label{sec:conclusions}

The main conclusion of the SuperCAM CO(3-2) survey can be summarized as follows: 

\begin{enumerate}
    \item We found a median value $R_{32}$=0.65$\pm$0.30 for the SMC.
    \item The $R_{32}$ values depend on the local environmental conditions of the region. For quiescent regions, $R_{32}$ $\lesssim 0.6$ were found. While for active star-forming regions, $R_{32}$ is between $\sim 0.7-1.0$.
    \item We identified $225$ CO($3-2$) clouds at 6 pc resolution in the SMC-Bar. The total luminosity in the $J=3-2$ transition is $(2.2 \pm 0.3)\times10^{4}$ \Lco\ within the mapped region.
    \item The CO(3-2) clouds follow similar scaling relations as the CO(2-1) clouds.  We found that the size-linewidth relation, for the CO($3-2$) clouds, is below the Milky Way clouds of similar size by a factor of 1.3. 
    \item Assuming that the CO(3-2) clouds are virialized, we determined a median value $\alpha_{\rm CO(3-2)} = 12.6_{-7}^{+10}$ \Xco\ for the total sample.
\end{enumerate}

\begin{acknowledgements}
      We thank the entire APEX crew for their continued support during the preparation of the SuperCAM visiting run, during the installation of the instrument, commissioning, and operation, and for general hospitality on site. H.P.S acknowledges partial financial support from a fellowship from Consejo Nacional de Investigaci\'on Cient\'ificas y T\'ecnicas (CONCET-Argentina) and partial support from ANID(CHILE) through FONDECYT grant No1190684. M.R. wishes to acknowledge support from ANID(CHILE) through FONDECYT grant No1190684 and ANID Basal FB210003. 

\end{acknowledgements}

%
\bibliographystyle{aa} 
\bibliography{biblio2} 
%

\begin{appendix} 

\section{RMS total map}
\label{app:rms_emission}

   \begin{figure}
   \centering
   \includegraphics[width=\linewidth]{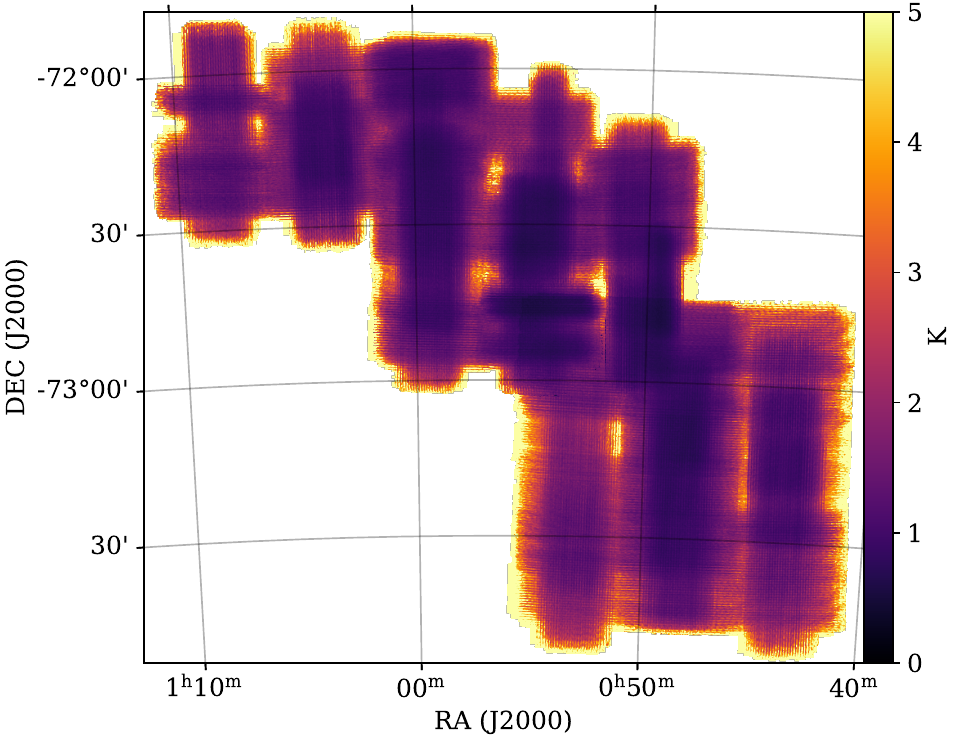}
      \caption{RMS map of the SuperCAM observation toward the SMC-Bar. Not considering the borders of this map, the $rms$ are between $0.1$ and $5$ K, with a median value of $\sim 1.0$ K.}
         \label{fig:smc_rms}
    \end{figure}

\section{$R_{32}$ distribution in Star Forming Regions}
\label{app:R32_distribution}
\begin{figure*}
\centering
\includegraphics[width = 0.7\linewidth]{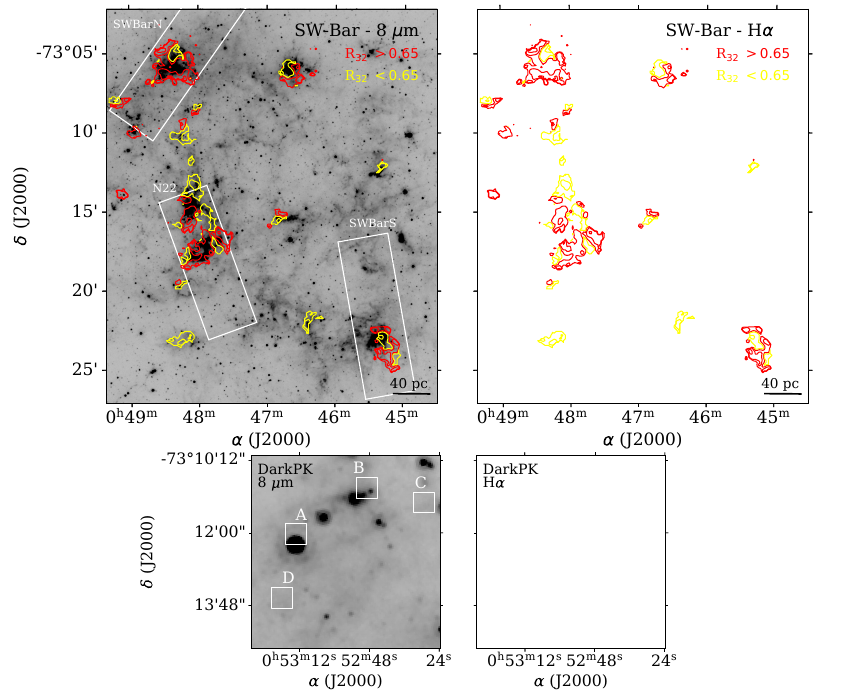}
\caption{Spatial distribution of $R_{32}$ (colored contours) overlapped to the Spitzer $8$ \mum\ (left panels) and H$\alpha$ emission (right panels) in grayscale for the SW-BAR and DarkPK regions. The yellow and red contours on the top panels correspond to $R_{32} = 0.2,~0.4,~0.6$ and $0.7, ~1.0,~2.0$, respectively. In the Bottom panels, the boxes A, B, C, and D indicate the position where $R_{32}$ was estimated (see Fig. \ref{fig:CO_lines}).}
\label{fig:SW_ratio_Halpha}
\end{figure*}

\begin{figure*}
\centering
\includegraphics[width = 0.7\linewidth]{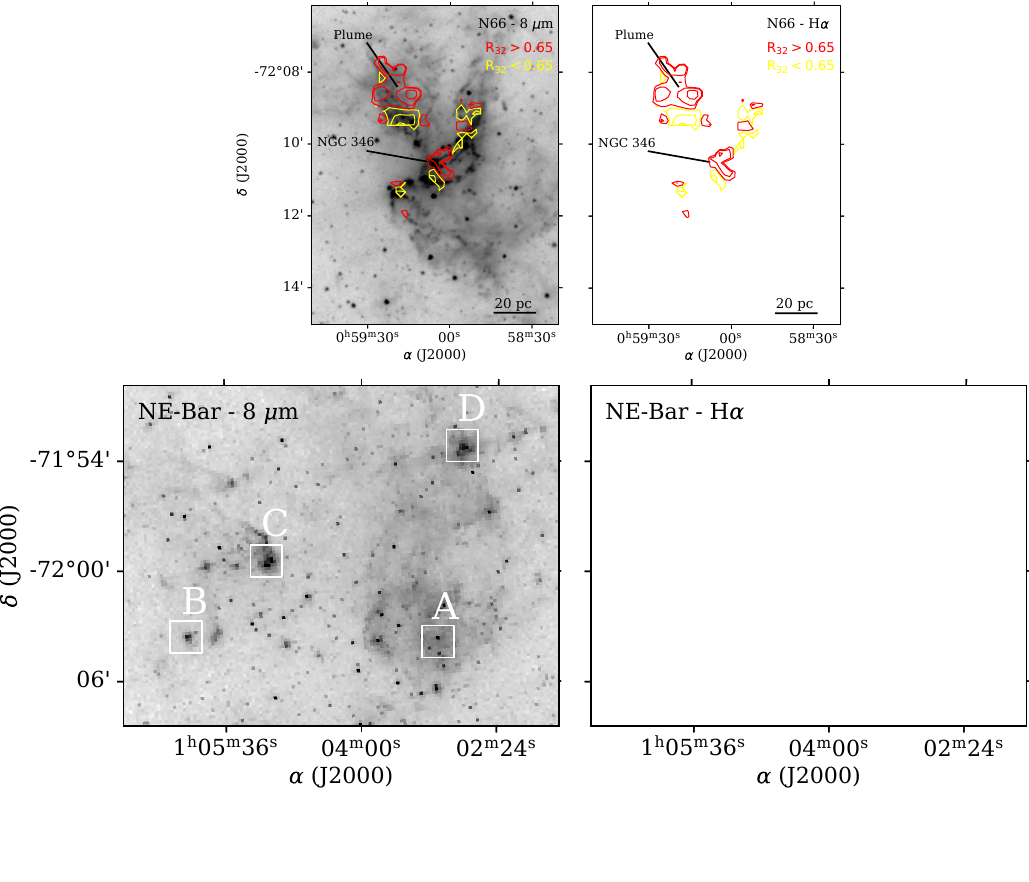}
\caption{Similar of Fig. \ref{fig:SW_ratio_Halpha} but for N66 and NE-Bar.}
\label{fig:NE_ratio_Halpha}
\end{figure*}
 
\end{appendix}

\end{document}